\documentclass{emulateapj}
\usepackage{placeins}
\usepackage{graphicx}

\shorttitle{Planets and disk gaps II} \shortauthors{Zhu et al.}

\newcommand\msun{\rm M_{\odot}}

\newcommand\msunyr{\rm M_{\odot}\,yr^{-1}}
\newcommand\be{\begin{equation}}
\newcommand\en{\end{equation}}
\newcommand\cm{\rm cm}

\newcommand\etal{{\rm et al}.\ }

\begin{document}

\title{ Dust Filtration by Planet-Induced Gap Edges: Implications for Transitional Disks}

\author{Zhaohuan Zhu\altaffilmark{1,2},
Richard P. Nelson\altaffilmark{3}, Ruobing Dong\altaffilmark{1}, 
Catherine Espaillat\altaffilmark{4}, and Lee Hartmann\altaffilmark{2}}

\altaffiltext{1}{Department of Astrophysical Sciences, 4 Ivy Lane, Peyton Hall,
Princeton University, Princeton, NJ 08544}
\altaffiltext{2}{Dept. of Astronomy, University of Michigan, 500
Church St., Ann Arbor, MI 48109}
\altaffiltext{3}{Astronomy Unit, Queen Mary University of London,
Mile End Road, London E1 4NS UK}
\altaffiltext{4}{Center for Astrophysics,
60 Garden St., Cambridge, MA 02138}

\email{zhzhu@astro.princeton.edu, r.p.nelson@qmul.ac.uk, rdong@astro.princeton.edu, cespaillat@cfa.harvard.
edu, lhartm@umich.edu }

\begin{abstract}
By carrying out two-dimensional two-fluid global simulations, we have studied the response of dust to 
gap formation by a single planet in the gaseous component of a protoplanetary disk - the so-called
``dust filtration'' mechanism.  We have found that a gap opened by a giant planet at 20 AU in a $\alpha$=0.01, $\dot{M}=10^{-8}\msunyr$ disk 
can effectively stop dust particles larger than 0.1 mm drifting inwards, leaving a 
sub-millimeter dust cavity/hole. However, smaller particles are difficult to
filter by a planet-induced gap due to 1) dust diffusion, and 2) a high gas accretion velocity at the gap edge.
Based on these simulations, an analytic model is derived to 
understand what size particles can be filtered by the planet-induced gap edge. We show that a dimensionless
parameter $T_{s}/\alpha$, which is the ratio between the dimensionless dust stopping time 
and the disk viscosity parameter, 
is important for the dust filtration process.
Finally, with our updated understanding of dust filtration, 
we have computed Monte-Carlo radiative transfer models with variable dust size distributions
 to generate the spectral energy 
distributions (SEDs) of disks with gaps.  By comparing with transitional disk observations 
(e.g. GM Aur), 
we have found that dust filtration alone has difficulties to deplete 
small particles sufficiently to explain the near-IR deficit of moderate $\dot{M}$ transitional disks, except
under some extreme circumstances.
The scenario of gap opening by multiple planets studied previously suffers the same difficulty.
One possible solution is by 
 invoking both dust filtration and dust growth in the inner disk. 
In this scenario,  a planet induced gap filters  large dust 
particles in the disk, and the remaining small dust particles passing to the inner disk can grow efficiently without 
 replenishment from fragmentation of large grains. Predictions for ALMA have also been made based on all these scenarios. 
 We conclude that dust filtration with planet(s) in the disk 
 is a promising mechanism to explain submm observations of transitional disks but it may need to be combined with
 other processes (e.g. dust growth) to explain the near-IR deficit of some systems. 

\end{abstract}

\keywords{accretion disks, stars: formation, stars: pre-main
sequence}

\section{Introduction}
The transitional and pre-transitional disks are protoplanetary disks around young stars
which exhibit strong dust emission at wavelengths $\gtrsim 10 \mu$m, while showing significantly
reduced fluxes relative to typical T Tauri disks at shorter wavelengths
(e.g., Strom \etal 1989,  Calvet \etal 2002, 2005; D'Alessio \etal 2005;
Espaillat \etal 2007,  2008). Pre-transitional disks still have
some infrared emission from warm, optically-thick dust near the star
(Espaillat \etal 2007, 2008, 2010), while transitional disks only have near- and 
mid-infrared emission from optically-thin
dust (Calvet \etal 2002, 2005; Espaillat \etal 2010). The depletion
of near- to mid-infrared emission is generally interpreted as being due to
evacuation of the disk interior to scales $\sim 5$ to $\sim 50$~AU
(Marsh \& Mahoney 1992; Calvet \etal 2002, 2005; Rice \etal 2003;
Schneider \etal 2003; Espaillat \etal 2007, 2008, 2010; Hughes \etal 2009),
an interpretation confirmed in some cases via direct sub-mm imaging
(e.g., Pietu \etal 2006; Brown \etal 2007, 2009; Hughes \etal 2009; Andrews \etal 2009, 2011).

These cavities need to be both  {\em large} and {\em deep} concerning the dust surface density
(e.g. Espaillat \etal 2010). Here {\em deep} means the gap is optically thin at mid-IR.
In the case of the transitional disks, the optically-thin region must extend from radii as
large as tens of AU all the way in to the central star, forming a dust hole.  Even the
pre-transitional disks, which display evidence of optically-thick dust emission
in the innermost regions, must have large disk gaps from $\sim$AU scales to tens of AU.
The requirement that the gaps/holes are optically thin implies that the
population of dust in sizes of order a micron or less must be extremely small, due 
to the large opacity of these small dust grains. 

Furthermore, many of 
these objects also exhibit gas accretion rates comparable 
to but slightly smaller than T Tauri disk accretion rates
($\sim 10^{-8}$$\msunyr$; Hartmann et al. 1998) onto
their central stars (e.g. Calvet \etal 2002, 2005; Espaillat \etal 2007, 2008;
Najita et al. 2007).  Maintaining this accretion requires a
significant mass reservoir in the inner disk. 
The conflict between the moderate disk accretion rate (which means a high gas surface density)
and the near-infrared deficit (which means a low dust density) puts strong constraints
on any theoretical attempts to explain these objects (Zhu \etal 2011).

In summary, there are three observational constraints for (pre)transitional disks: 
1) A wide gap/hole extending over one order of magnitude in radii, 
2) A deep gap/hole that is optically thin, and 3) a moderate
gas accretion rate.

Gap opening by multiple giant planets (Zhu \etal 2011, Dodson-Robinson 
\& Salyk 2011) has been proposed to
explain the wide gaps in (pre)transitional disks. However, Zhu \etal (2011) also pointed
out that even with four giant planets, the opened gap is not deep enough to explain transitional
disks, unless there are massive giant planets close to the central star.  Zhu \etal (2011)
concluded that dust depletion/growth is still required in the inner disk to be consistent with
observations. On the other hand, Dodson-Robinson \& Salyk (2011) speculated that dust may be significantly
confined in the spiral density wakes or streamers.
Both of these works imply that dust dynamics may need to be considered along with the gas dynamics.

By considering dust dynamics independently, one promising mechanism to explain 
these wide and deep (dust) gaps/holes is dust filtration by the
gap outer edge opened by planet(s) (Paardekooper \& Mellema 2006, Rice \etal 2006). This mechanism
is intrinsically due to the dust 
being
a pressureless system, while the gas orbital dynamics is affected by the gas pressure. 
Dust particles orbit the central star at the Keplerian speed, but the gas rotates 
 slightly faster or slower
than this depending on the radial pressure gradient. In this case the dust particles
feel either a headwind or tailwind from the gas particles and can lose or gain angular momentum
to the gas and start drifting radially in the disk. The net effect is that  dust
particles tend to drift toward gas pressure maxima. Thus, at the outer edge of a 
planet-induced gap where the radial density/pressure gradient is positive, 
dust particles drift outwards, possibly overcoming their coupling to the inward gas accretion process. 
Dust particles 
will then stay at the gap outer edge while the gas flows through the gap. 
This process is called 'dust filtration' (Rice \etal 2006), and
it depletes the disk interior to the radius of the planet-induced gap of dust, forming
a dust-depleted inner cavity. 

However, this filtration process sensitively depends on the pressure gradient
at the planet-induced gap outer edge and it only operates for large particles whose drift velocities are significant. 
Small
particles strongly coupled to the gas flow can still penetrate the planet-induced
 gap to the inner disk. More importantly, dust diffusion may significantly reduces the filtration efficiency 
 even for the big particles (Ward 2009).
Thus it is crucial to understand what sized particles can be filtered by a realistic planet-induced gap.
In this paper, we will show only 0.1 mm particles and above can be filtered by a realistic gap if the disk
viscosity parameter $\alpha$=0.01 and accretion rate $\dot{M}=10^{-8}\msunyr$. This
size limit may decrease to 0.01 mm if the planet is very massive and the disk is less turbulent accreting at a low accretion rate.
 Although this
can explain the cavity from submm observations  of transitional disks, it cannot explain the near-IR
deficit of these disks, given that significant small-dust is still present  in the inner disk. 
However, on the other hand, if dust filtration is combined with dust growth in the disk, the near-IR deficit of transitional disks can be very well reproduced. 
 
In \S 2, we introduce the two-dimensional two-fluid simulations. In \S 3, we construct a
simpler one dimensional model. Our results are presented in \S 4. A short discussion is presented in \S 5. Various
scenarios trying to explain transitional disks are tested with the Monte-Carlo radiative transfer model in \S 6, 
and conclusions are
drawn in \S 7.

\section{2D 2-fluid simulations}
In this section, we will first introduce our 2-fluid numerical algorithms, and then introduce our special inner boundary condition
to study long timescale disk evolution. After a short discussion on the numerical challenging of simulating dust fluid, we set-up our
gas/dust 2-fluid simulations. 
\subsection{Numerical Algorithms}
Gaseous and dust components of the disk are simulated separately in our simulations.
\subsubsection{Gas component}
The gas disk evolution is simulated with FARGO (Masset 2000), a two-dimensional hydrodynamic code
which utilizes a fixed grid in cylindrical polar coordinates (R, $\phi$). FARGO uses finite differences to approximate
derivatives, and the evolution equations are divided into source and transport steps, similar to those
of ZEUS (Stone \etal 1992) . However, an orbital advection scheme has
been incorporated which reduces the numerical diffusivity and significantly
increases the allowable timestep as limited by the
Courant-Friedrichs-Lewy (CFL) condition. 
Thus FARGO enables us to study the interaction between the disk and embedded
planets  over a full viscous timescale for disks. The set-up of the gaseous disk and the planet
is discussed in \S 2.4.1.

\subsubsection{Dust component}
Beyond the gas component, we have implemented an additional fluid in FARGO to simulate the dust's response to the gas. The dust is treated as a low pressure fluid and couples with the gas via drag terms. No feedback from the dust on the gas is simulated, since dust filtration normally takes place when the gas density dominates the dust density.  However, this assumption
may be violated under some circumstances as discussed in \S 5.4.  The drift terms are added as an additional source step for the dust fluid
\begin{equation}
\frac{\partial v_{r,d}}{\partial t}=-\frac{v_{r,d}-v_{r,g}}{t_{s}}\,,\label{eq:vrd}
\end{equation}
\begin{equation}
\frac{\partial v_{\theta,d}}{\partial t}=-\frac{v_{\theta,d}-v_{\theta,g}}{t_{s}}\,,\label{eq:vtd}
\end{equation}
where all other terms/steps are the same as the gaseous fluid (Stone \& Norman 1992). Since we will focus on dust particles with radii smaller than 1 mm\footnote{With our disk parameters,
the mean free path of the molecule is 0.1 mm at 0.1 AU and 5 m at 10 AU, which is larger than the dust size. At the very inner disk, 
where the mean free path is small and the particles are no longer in the Epstein regime, the viscous velocity dominates the drift velocity so that whether the particles are in the Epstein regime is no longer important.},
these particles are in the Epstein regime (Whipple 1972, Weidenschilling 1977), so that
 the dust stopping time (Takeuchi 
\& Lin, 2002) is
\begin{equation}
t_{s}=\frac{ s \rho_{p} }{\rho_{g}v_{T}}\,,
\end{equation}
where $\rho_{g}$ is the gas density, s is the dust particle radius, $\rho_{p}$ is the dust particle density (we chose $\rho_{p}$=1 g cm$^{-3}$), v$_{T}$=$\sqrt{8/\pi}c_{s}$, and $c_{s}$ is the gas sound speed. Considering the mean disk density is $\rho_{g}=\Sigma_{g}/\sqrt{2\pi} H$, it can also be written as
\begin{equation}
t_{s}=\frac{\pi s \rho_{p} }{2\Sigma_{g}\Omega}\,.
\end{equation}
 In a dimensionless form the stopping time can be written as 
\begin{equation}
T_{s}=t_{s}\Omega
\end{equation}
where $\Omega$ is the Keplerian angular velocity.

When the dust stopping time is far shorter than the hydro time step, the drift terms become stiff. With an explicit method,  the time step $\Delta$t needs to be smaller than both the stopping time t$_{s}$, and the hydro time step constraint $\delta$t  in order to be numerically stable.  However, t$_{s}$ is very small with micron size particles, thus an implicit method is desired. Considering that the ZEUS/FARGO scheme is first order accurate in time, the first order implicit scheme is
\begin{equation}
 v^{n+1}_{r,d}=v^{n}_{r,d}-\frac{v^{n}_{r,d}-v^{n}_{r,g}}{t_{s}+\Delta t}\Delta t\,,\label{eq:imp}
\end{equation}
where $n+1$ denotes the quantities at the new time step.
Since the drift term is stiff, it may be helpful if we go to a higher order scheme. It turns out that the second order scheme just requires replacing the $\Delta t$ in the denominator with $\Delta t/2$. 

Although the implicit method ensures a stable scheme no matter what the time step we use,  Equation \ref{eq:imp} suggests it merely adds $\Delta t$  to the stopping time. The new equivalent stopping time is $\Delta t + t_{s}$ which can be significantly larger than the real $t_{s}$ if the numerical time step is much larger than $t_{s}$. Thus in order to accurately study the dust drift which is controlled by $t_{s}$, the hydro time step needs to be comparable to $t_{s}$. By numerical testing we found that the radial drift velocity is close to the theoretical value  only if the time step is close to $t_{s}$. We tried different schemes as mentioned above, but they differ little. Thus, to simulate the dynamics of dust particles much smaller than 1 mm correctly, the time step is set by the dust stopping time and is inversely proportional to the particle size. 
Simulating dust drift for 0.01 mm particles is thus very computationally expensive since it is 100 times 
more expensive than simulating 1 mm particles.   

 On the other hand, we are not interested in dust dynamics happening on the dust stopping timescale. We are only 
 interested in the dust's final response to the gas flow which changes on a much longer timescale than the stopping time. 
 In other words, we can use the dust's terminal velocity to represent its final response to the gas. 
 This is introduced as the ``Short Friction Time Approximation'' (SFT) in  
Johansen \& Klahr (2005), where the dust velocity is related with the gas velocity as
\begin{equation}
\bold{v_{d}}=\bold{v_{g}}+t_{s}\frac{\nabla P}{\rho}\,.\label{eq:eqSFT}
\end{equation}
This approximation is valid only if t$_{s}$ is much smaller than the dynamical timescale of the gas flow. More accurately, t$_{s}$
needs to be smaller than the hydrodynamic time step. In our simulation set-up, even with 1mm particles, 
t$_{s}$ is at least one order of magnitude
smaller than the hydrodynamic time step.  In the Appendix we compare the SFT approximation 
with the two-fluid simulation and show good agreement between them.

Furthermore, dust can diffuse in the gaseous disk due to turbulence.
In this work dust diffusion  is  
implemented in the operator split fashion in the source step for the dust fluid (Clarke 
\& Pringle 1988)
\begin{equation}
\frac{\partial \Sigma_{d}}{\partial t}=\nabla \cdot \left(D\Sigma_{g}\nabla\left(\frac{\Sigma_{d}}{\Sigma_{g}}\right)\right)\,,\label{eq:diffusion}
\end{equation}
where D is the turbulent diffusivity which relates with the turbulent viscosity $\nu$ through
\begin{equation}
D=\frac{\nu}{Sc}\,,
\end{equation}
where $Sc$ is the Schmidt number defined as the ratio between the total accretion stress 
and particle mass diffusivity. Note that $\nu$
here is the viscosity of the `gaseous' disk, representing the efficiency 
of the angular momentum
transport experienced by the gas disk due to disk turbulence. 
When the disk orbital time (which is close to the turbulent eddy turnover time) is much larger 
than the dust stopping time ( which is always true for particles smaller than $\sim 1$~mm), $Sc\sim$1 (Johansen \& Klahr 2005, 
Carballido et al. 2011) and dust will  not settle significantly so that 2-D approximation is better justified. 
The above formula for dust diffusion has been confirmed by
both analytic work and numerical simulations including both MRI turbulence and 
 particle dynamics under the circumstances that the background gas surface density
 is uniform
  (Youdin\& Lithwick 2007, Carballido et al. 2011). 
In \S 4, we discuss how diffusion plays a significant role in the dust filtration process.

\subsection{Inner Boundary Conditions}
For the gaseous fluid, as discussed by Crida, Morbidelli, \& Masset (2007),
a standard open inner boundary condition (Stone \etal 1992) in a fixed 2D grid can
produce an unphysically rapid depletion of material through
the inner boundary in the presence of the planets.
There are two reasons for this.  First, due to waves excited by the planet,
the gas in the disk can have periodic inward and
outward radial velocities larger than the net viscous velocity of accreting
material.
Thus, with the normal open boundary, material can
flow inward while there is no compensating outflow allowed.
Second, the orbit of the gas at the inner boundary is not circular
due to the gravitational potential of the planets;
again, as material cannot pass back out
through the inner boundary, rapid depletion of the inner
disk material is
enhanced.
As we are interested in the amount of gas depletion in the disk inward of the
planet-induced gap(s) over substantial evolutionary timescales,
it is important to avoid or minimize this unphysical mass depletion.

Crida \etal (2007) were able to ameliorate this problem by surrounding the
2D grid by extended 1D grids (FARGO2D1D, see their Figure 5).  We follow Pierens \& Nelson
(2008), who found reasonable agreement with the Crida \etal results while
using a 2D grid only by limiting the inflow velocities at the inner boundary
to be no more than 3 times larger
than the viscous radial velocity in a steady state,
\begin{equation}
v_{rs} = - \frac{3 \nu_{in}}{2R_{in}} \,,\label{eq:vrs}
\end{equation}
where $\nu_{in}$ and $R_{in}$ are the viscosity and radius at
the inner boundary. This scheme shows good agreement
with the analytic estimate and the results from FARGO2D1D model (Pierens \& Nelson 2008, Zhu \etal 2011). 

For the dust fluid, we adopt a similar approach by limiting the radial velocity 
of the dust fluid to be no more than 3 times larger than the dust drift speed in a viscous disk,
\begin{equation}
v_{rs,d}=\frac{-(3 \nu_{in}/2R_{in})T_{s}^{-1} - \eta v_{K}}{T_{s}+T_{s}^{-1}}\,, \label{eq:vd}
\end{equation}
where $\eta$ is the ratio between the pressure gradient and gravitational force
$\eta$ = $-(R\Omega^{2}\rho_{g})^{-1}\partial P/\partial R$, and it is equal to 
3/2(H/R)$^{2}$ in our set-up below.

\subsection{Fluids with and without pressure}
A zero pressure fluid means zero scale height, which implies any velocity disturbance in the disk will quickly sharpen and shock, known as the delta shock \footnote{If a delta shock forms in the disk, the fluid treatment
needs to be replaced by a particle treatment since particles will
cross orbits changing their distribution function in phase space.}.
In reality, a shock won't form in the particle fluid if there is a strong
coupling between gas and dust . 
Numerically due to the inability to simulate coupling on scales smaller than the grid,
a zigzag shaped density profile forms grid by grid in our simulations
and we rely on the artificial viscosity to stabilize the shock. 
To minimize this effect, a small pressure is applied to the dust fluid here,
 which makes the scheme robust but it also introduces limitations as described below.
 
After numerical tests, we chose the dust scale height H$_{d}$/R=0.0044 for the dust fluid, which is one order of magnitude smaller than the gaseous fluid (H/R=0.044).
H$_{d}$ is close to our grid spacing, while significantly smaller than the gas disk scale height. Since dust drift is intrinsically due
to the gaseous disk pressure gradient, adding 10$\%$ pressure to the dust fluid means the drift speed is only affected by 10$\%$. But
if the dust fluid has a sudden density change (e.g. 10 times steeper than the gaseous fluid) the effect of the small pressure can be amplified and
becomes erroneous.

\subsection{Model set-up}
\subsubsection{Gas component}
Similar to Zhu \etal (2011), we assume a central stellar mass of $1 \msun$ and a fully viscous disk.
We further assume a radial temperature distribution
$ T = 221 (R/{\rm AU})^{-1/2}$~K, which is roughly consistent with typical T Tauri
disks in which irradiation from the central star dominates the disk temperature
distribution (e.g., D'Alessio \etal 2001). The disk is vertically isothermal.
The adopted radial temperature distribution corresponds to an implicit ratio
of disk scale height to cylindrical radius $H/R = 0.029 (R/AU)^{0.25}$.
This differs from the $H/R =$~constant assumption used in many previous
simulations, which implies a temperature distribution
$T \propto R^{-1}$, which is inconsistent with observations. Consequently,
our assumed temperature distribution with a constant
viscosity parameter $\alpha$ ($\nu =\alpha c_{s}^{2}/\Omega$, where $\nu$ is the
kinematic viscosity, $c_s$ is the sound speed, $\Omega$ is the angular velocity)
leads to a steady-disk surface density distribution
$\Sigma \propto R^{-1}$ instead of the $\Sigma \propto R^{-1/2}$ which would
result from either assuming both $H/R $ and $\dot{M}$ are constant, or both $\nu$ and viscous torque (-2$\pi R\Sigma\nu R^{2}d\Omega/dR$) are constant.
This makes a significant
difference in the innermost disk surface densities, and thus the
implied inner disk optical depths in our models will be larger.

We set $\alpha = 0.01$ for the standard cases.
Given our assumed disk temperature distribution,
we set the initial disk surface density to be
$\Sigma$=178 (R/AU)$^{-1}$exp(-R/100 AU)  g cm$^{-2}$
from $R \sim 8 - 300$~AU, which yields a steady
disk solution with an accretion rate $\dot{M} \sim 10^{-8} \msunyr$,
typical of T Tauri disks (Gullbring \etal 1998; Hartmann \etal 1998).

\subsubsection{Dust component}
 The dust surface density is assumed to be 0.01 of the gas surface density initially. With this set up, the dust stopping time 
$t_{s} \propto 1/(\Sigma_{g} \Omega) \propto R^{5/2}$. Motivated by (pre-)transitional disk observations, we place the planet at 20 AU. 

We have carried out two-fluid simulations, with a 1 M$_{J}$ planet in the gaseous and dust disk (1 mm particles) to 2.5$\times$10$^{4}$ years, using 
two methods for the dust fluid: 1) self-consistently solving the dust fluid equations together with the gas equations, or
 2) using the SFT approximation (Equation \ref{eq:eqSFT}). In the following we will refer to the former as two-fluid 
simulations, and the second method as the SFT approximation (although the second method is also a two-fluid method).
 Good agreement has been found 
between these two methods (see Appendix). Since the SFT approximation is not limited by the dust stopping time, we use this approximation for 
the dust component in all the other runs.

Finally, gaseous and dust disks including three different size particles (1, 0.1, and 0.03 mm) and three different mass planets (1, 3, 6 M$_{J}$) have been simulated, respectively. 
All the runs are summarized in Tabel 1 and discussed in \S 4.3. 

\section{1 D simulation}
Because the dust filtration process is mainly determined by the planet-induced gap structure which
is quite axisymmetric except in the region close to the planet,
 it is useful to construct simpler 
azimuthally averaged 1-D models; by
comparing 1-D models with 2-D simulations, we can separate effects due to axisymmetric and non-axisymmetric features. For axisymmetric
flows, the dust surface density evolution is governed by the continuity equation (Takeuchi 
\& Lin 2005)
\begin{equation}
\frac{\partial \Sigma_{d}}{\partial t}+\frac{1}{R}\frac{\partial}{\partial R}[R(F_{diff}+\Sigma_{d}v_{d})]=0\,,\label{eq:sigd}
\end{equation}
where $F_{diff}$ is the dust mass flux due to diffusion,
\begin{equation}
F_{diff}=-D\Sigma_{g}\frac{\partial}{\partial R}\left(\frac{\Sigma_{d}}{\Sigma_{g}}\right)\,,
\end{equation}
 and $v_{d}$ is the dust radial velocity in the rest frame due to the gas-drag
\begin{equation}
v_{d}=\frac{v_{g}T_{s}^{-1} - \eta v_{K}}{T_{s}+T_{s}^{-1}}\,,\label{eq:driftv}
\end{equation} 
where $T_{s}$, $\eta$ are defined above. 
When $T_{s} \ll 1$ the above equation can be approximated by
\begin{equation}
v_{d}=v_{g}-\eta v_{K}T_{s}\,.\label{eq:driftv2}
\end{equation}
We can also incorporate the diffusion term in Equation \ref{eq:sigd} into the the dust velocity so that the equivalent total dust velocity is
\begin{equation}
v_{d,t}=v_{g}-\eta v_{K}T_{s}-\frac{D}{R}\frac{d \rm{ln}(\Sigma_{d}/\Sigma_{g})}{d \rm{ln} R}\,\label{eq:driftv3}\,,
\end{equation}
where the first term on the right hand side
is the dust radial velocity induced by the gas radial velocity (normally it is negative due to the accretion process), 
the second term is the dust drift velocity with respect to the gas 
due to the gas pressure gradient (at the gap outer edge $\eta$ is negative),
and the last term is the dust velocity due to dust diffusion.
The addition of the first two terms is the dust drift velocity in the rest frame without diffusion.

 The only quantities
which cannot be calculated in the 1-D models are the gaseous disk surface density  with a gap opened by a planet
 ($\Sigma_{g}$, which is
needed to calculate $\eta$) and the gas radial
velocity $v_{g}$. Here, we adopt the gaseous disk surface density ($\Sigma_{g}$) from 2-D simulations, and
the initial dust surface density is chosen as 1/100th of the gas surface density.
We also
assume the gas radial mass flux at each radius is a constant (equal to $\dot{M}$), giving
\begin{equation}
v_{g}=\frac{\dot{M}}{2\pi R \Sigma_{g}}\,.\label{eq:radialgas}
\end{equation}  
This implies the gas velocity can be quite large deep inside the gap where the gas surface
density is very low. This high velocity has significant impact on the dust drift as discussed in 
Rice \etal (2006) and \S 4.3.
In 1-D, the radial velocity advection is ignored and the azimuthal velocity is assumed to be Keplerian.
Thus we can simulate the pressureless fluid without worrying about shock formation. Furthermore,
in our set-up, $v_{g}$ and $\Sigma_{g}$ are fixed, so that $v_{d}$ is also fixed and the part of Equation \ref{eq:sigd} without diffusion
 is a linear equation for $\Sigma_{d}$. Due to the simplicity of the 1-D model, it not only extracts the essential physics but 
 can also serve as a sanity check for 2-D simulations.
 
\section{Results}
Before we present any results, we want to emphasize that 
 Equation
\ref{eq:driftv3} guides all our discussions below. Dust filtration
happens when Equation \ref{eq:driftv3} (or Equation \ref{eq:driftv2} if there is no dust diffusion) becomes positive so that
the total dust velocity is outwards. 
In the following, we will first present the simplest case without dust diffusion and then we will present results with
dust diffusion considered. 

\subsection{Dust Filtration without Diffusion}
\begin{figure}
\epsscale{.80} \plotone{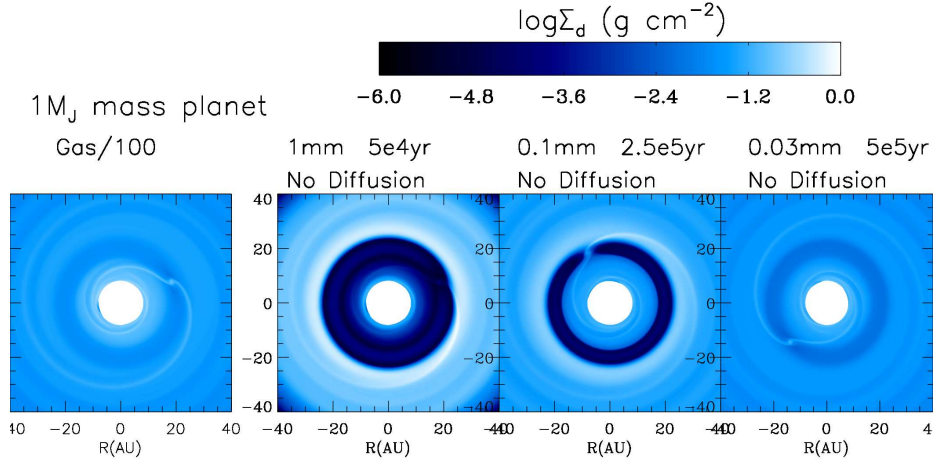} \caption{The disk gas (left panel) and different sized dust (right panels) surface density contours
if a 1 $M_{J}$ planet is at 20 AU. The gas surface density is divided by 100 to scale with the dust surface densities.
Three different sized dust  has been shown in the right panels. Dust diffusion has been turned off in these simulations.
For 1 mm particles, the contour shows the surface density at 5$\times$10$^{4}$ yr, 
which is longer than its radial drift timescale.
For 0.1 mm particles, the surface density shown is at 2.5$\times 10^{5}$ yr, which is even longer than the disk
viscous timescale at 20 AU. For 0.03 mm particles,   the surface density shown is at 5$\times 10^{5}$ yr.
Clearly, 1 mm particles are filtered by the planet-induced gap and no inner 1 mm dust disk exists. 
Particles smaller than 1 mm can penetrate the gap freely and form an inner dusty disk. 
} \label{fig:fig1}
\end{figure}

Although a 1 M$_{J}$ planet can only open a shallow gap in the gas disk in our set-up, 
1 mm dust particles can be effectively trapped at the gap outer edge.
As shown in the left panel of Figure \ref{fig:fig1}, the gaseous gap is barely apparent and the depth of the gap is half of the unperturbed disk. 
 However,  if dust diffusion is ignored, such a shallow gaseous gap can effectively stop 1 mm particles
at the gap outer edge due to the dust drift. Then without replenishment, 1 mm dust particles in the inner disk 
quickly move inward to the central star, leaving a large mm dust cavity/hole in the disk  (the second to left panel of Figure \ref{fig:fig1}).
The cavity is highly depleted by four orders of magnitude. Thus the planetary wake is less apparent 
within the cavity, but it can still be
seen outside the cavity. One feature which can be seen but is not very apparent in the dust image is the dust ring in the horseshoe region (the ring can also
be observed in Figure \ref{fig:fig4}). The ring forms because
the gas pressure has a local maximum in the center of the horseshoe region, effectively trapping dust particles.

On the other hand, smaller particles (0.1 mm, the second to right panel) can still penetrate the planet-induced gap to the inner disk.
This is because, without dust diffusion, the dust outward drift velocity with respect to the gas (the second term in Eq \ref{eq:driftv2}) decreases as $T_{s}$ decreases and becomes comparable to the 
inward gas speed (the first term in Eq \ref{eq:driftv2}) so that the total dust velocity can be inward towards the central star. 
Since  $T_{s}$
is proportional to the particle size, it means smaller particles may not have enough outward drift to counteract the global accretion speed
and they will be carried inwards across the gap by the gas, although at  a small speed.
This small speed, which implies a low dust accretion rate, can lead to the particles piling 
up at the outer edge of the gap. Thus the dust concentration
is slightly increased at the outer edge of the gap, which makes the gap look deeper. 
But generally, the spatial structure of the small dust particles looks similar to that of 
the gaseous disk without a highly depleted cavity/hole. 

For smaller particles (eg. 0.03 mm in the right-most panel of Figure \ref{fig:fig1}), 
$T_{s}$ decreases, and the dust disk behaves more similarly to the gaseous disk. 
In the extreme limit that dust particles are very small so that the drift velocity is negligible compared with the accretion velocity,
the structure of the dust disk looks exactly like that of the gaseous disk.

\subsection{Dust Filtration with Diffusion}
The dust filtration process can be significantly hindered by dust diffusion,
 as originally
pointed out by Ward (2009).

Dust diffusion tries to smooth any density feature of the dust relative to the gas. Thus, at the gap outer edge, 
dust tries to diffuse inwards, leading to an additional inward dust speed (adding the third term in Equation \ref{eq:driftv3}), which lowers the outward
drift speed. The diffusion velocity (Eq \ref{eq:sigd}) is 
\begin{equation}
v_{diff}=-\frac{D}{R}\frac{d \rm{ln}(\Sigma_{d}/\Sigma_{g})}{d \rm{ln} R}\,.\label{eq:vdif}
\end{equation} 
Unlike the gas diffusion velocity
\begin{equation}
v_{g}=-\frac{3}{R^{1/2}\Sigma_{g}}\frac{d}{d R}\left(R^{1/2}\nu\Sigma_{g}\right)\,,
\end{equation}
which depends on the gas surface density,
the dust diffusion velocity depends on the dust concentration relative
to the gas rather than on its absolute abundance.

At the beginning of the simulation, the abundances of dust and gas do not vary much so that 
the concentration of dust
$\Sigma_{d}/\Sigma_{g} \sim 0.01$ everywhere and dust diffusion can be ignored. However, 
as shown in the last section, the dust quickly concentrates at the gap outer edge and 
dust diffusion can play a significant role. The timescale
for dust diffusion becoming important is the dust radial drift timescale, assuming the gaseous
disk is in steady state,
\begin{equation}
\tau_{dust}=\frac{R_{gap}}{v_{d}}
\end{equation}
where R$_{gap}$ is the position of the planet-induced gap and $v_{d}$ is defined in Equation \ref{eq:driftv}.
With our disk parameters ($\dot{M}$=10$^{-8} \msunyr$, $\alpha$=0.01, R$_{gap}$=20 AU),
$\tau_{dust}$=9$\times$10$^{4}$ yr for 1 mm particles, 2$\times$10$^{5}$ yr (the gas viscous timescale) for 0.1 mm 
and smaller particles.
Thus, to study dust filtration, we have to evolve the disk long enough. For small
dust particles ($\le$0.1 mm), it means to simulate the disk evolution at the disk viscous timescale at the position of the gap
( $\sim$2$\times$10$^{5}$yr ). In this work, we managed to carry out such simulations by 
using the SFT approximation. 

\begin{figure}
\epsscale{.80} \plotone{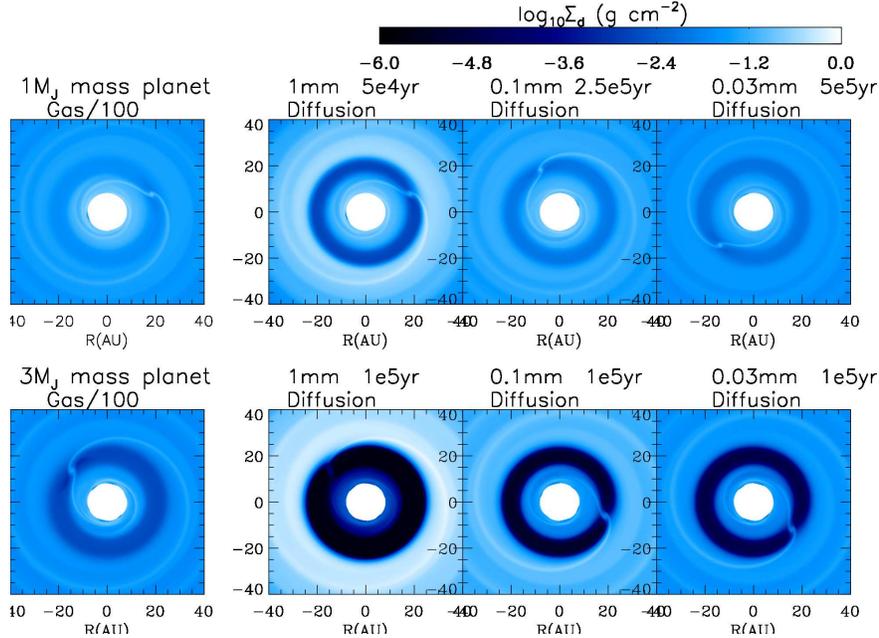} \caption{Upper panels: similar to Figure \ref{fig:fig1} but with dust 
diffusion due to disk turbulence included. 
Clearly dust diffusion leads to less efficient dust filtration. Bottom panels: Similar to the upper panel but with 
a 3 $M_{J}$ planet in the disk. The inner disk for 1 mm dust is depleted, while for 0.1 mm and 0.03 mm dust the
dust inner disk is still present, although the gap itself becomes deeper. 
Since the gaps in the bottom panels are very deep, the color bar is from -10 to 0 for the 1 mm case, -7 to 0 for
the 0.1 mm case, and -6 to 0 for the 0.03 mm case.
} \label{fig:fig2}
\end{figure}

Figure \ref{fig:fig2} shows that dust diffusion significantly hinders dust filtration. Compared with the right panels
of Figure \ref{fig:fig1}, the dust gap is much shallower and the dust distribution is a lot smoother. 1 mm dust
particles can still penetrate the gap opened by a 1 M$_{J}$ planet, unlike the case without diffusion. 
However, dust diffusion cannot stop dust filtration indefinitely. If the gap is steeper (e.g. a gap opened by a 3 M$_{J}$
planet, the lower panels of Figure \ref{fig:fig2}), 1 mm dust particles can still be filtered by the gap,
which will be discussed below.

\subsubsection{Varying Gap Stucture and Dust Sizes}
Figure \ref{fig:fig4} shows both gas and dust surface densitiy profiles for various planets
in the disk (1, 3, 6 M$_{J}$ from top to bottom) with and without dust diffusion (right and left panels, respectively).
The solid curves are the gas surface densities divided by 100. At the end of the simulation, 
if the dust surface density is depleted by more than 1000 at the inner boundary ($\>$3 orders of magnitude 
smaller than the solid curve), we say the dust is significantly depleted and filtered.  

The top panels show the disk surface densities with a 1 M$_{J}$ planet at 20 AU. As discussed above, 
the gaseous gap is quite shallow,
approximately half of the unperturbed disk surface density. Without dust diffusion, such a shallow gap is capable of
filtering 1 mm dust particles, but unable to filter smaller particles. However, with dust diffusion, all the dust particle sizes
we considered ($0.03 \le s \le 1$ mm) can pass through this 1 M$_{J}$ planet-induced gap. 
Dust diffusion also changes the dust distribution in the outer disk: 
dust can diffuse outwards, slowing down the shrinkage of the dust disk and making the
decline of the dust surface density with increasing radius more gradual. 

The middle panels show the disk surface densities with a 3 M$_{J}$ planet at 20 AU. The gaseous gap 
is one order of magnitude deep. Without dust diffusion, both 1 mm and 0.1 mm particles are filtered, while
0.03 mm particles can pass through. With dust diffusion, 1 mm particles are filtered, while 0.1mm and 0.03 mm
particles pass through.

The lower panels show the disk surface densities with a 6 M$_{J}$ planet at 20 AU. The gaseous gap
is almost two orders of magnitude deep. Without dust diffusion, all kinds of particles are filtered. With
dust diffusion, 1 mm and 0.1 mm particles are filtered. 

To emphasize the effect of dust diffusion we use the simplest 1-D model to show various components
of the dust velocity  in  Figure \ref{fig:fig5} for 1 mm particles in a gaseous disk with 1 M$_{J}$ mass planet. 
The left panel shows the case without dust diffusion,
while the right panel includes dust diffusion.
The gas velocity is the green curve. 
The total dust velocity is the solid black curve, 
including the dust drift velocity in the rest frame (the addition of the first and second terms of Equation \ref{eq:driftv3}, dotted curve) 
and the diffusion velocity (the third term of Equation \ref{eq:driftv3}, dashed curve). As shown in this 
figure, dust drift is fully capable of filtering the 1mm particles (the dust drift velocity in the rest frame is positive at the gap outer edge).
However dust diffusion adds an additional negative velocity component. The net effect is that 
the total dust velocity is still inwards (dust filtration fails) when dust diffusion is considered.

In summary, dust diffusion hinders the dust filtration process. 
1 mm-sized particles will be filtered by a gap one order of magnitude deep and 
particles with sizes $\ge 0.1$ mm will be filtered by a gap that is two orders of magnitude deep. 

\subsection{High Gas Velocity at the Gap Edge}
Another effect hindering filtration is the high gas velocity at the gap edge.
Although this effect is self-consistently treated in simulations, it needs
special attention in analytical approaches as 
 noted in Rice \etal (2006). We have identified and 
 verified this effect in our simulations.

Considering the gas accretion rate is constant across the planet-induced gap (if we allow the planet to accrete, this is 
still true at the outer edge of the gap), 
the radial velocity has to increase in the gap since the disk surface density decreases in the gap (according to
 $\dot{M}=2\pi R\Sigma_{g} v_{g}$).  With a two orders
of magnitude deep gaseous gap, the gas velocity can be amplified by two orders of magnitude within the gap. Thus
the first term of  Equation \ref{eq:driftv3} is significantly increased. Please note that this argument is built on
the assumption that 
the velocity is axisymmetric inside the planet-induced gap. Although this assumption is incorrect 
deep inside the gap around the horseshoe orbits (see Appendix), the flow is quite axisymmetric at the edge 
of the gap where dust filtration takes place. 
We have checked 2-D simulations directly and confirmed that
the radial velocity increases by a factor that is close to the gaseous gap depletion factor. This effect becomes increasingly important
for a deeper gaseous gap. 

To illustrate this effect, we again use the simplest 1-D model to show various components
of the dust velocity, as shown in Figure \ref{fig:fig6}. Even with
dust diffusion considered, by ignoring the amplification of the gas radial velocity across the gap (green curve
in the left panel is flat), 0.03 mm particles will be filtered in the 6 M$_{J}$ gap. However, considering
the gas velocity is amplified (the right panel), 0.03 mm particles can pass through the gap, consistent with the results from
2-D simulations.

\subsection{What size particles will be filtered? Analytical approach}
Although  our simulations have suggested that
1 mm particles will be filtered by a gap one order of magnitude deep and $\ge$0.1 mm particles will
be filtered by a gap that is two orders of magnitude deep, it would be
insightful to have a simple analytic model to show how the critical filtration particle size
depends on the disk parameters.

Both dust diffusion and high
gas velocity at the gap edge are important as pointed out above. Thus we will
consider them separately and then combine their effects.

First, we will assume the radial gas flow velocity is zero, and only consider the dust diffusion process to balance the dust drift. We are trying to 
find the marginal state between the dust being filtered and drifting inwards. This 
marginal state can be derived by assuming the diffusion velocity Equation \ref{eq:vdif}  balances the drift velocity 
\begin{equation}
-\nu\frac{d \ln(\Sigma_{d}/\Sigma_{g})}{d R}=\frac{\eta V_{K}}{T_{s}+T_{s}^{-1}}\,.\label{eq:nueta}
\end{equation}
Since $T_{s}\ll$1, and plugging in $\eta=-(R\Omega^{2}\Sigma_{g})^{-1} \partial P/\partial R$
and $\nu=\alpha c_{s}^{2}/\Omega$, we derive
\begin{equation}
\frac{d \ln(\Sigma_{d}/\Sigma_{g})}{dR}=\frac{T_{s}}{\alpha}\frac{\partial \ln\Sigma_{g}}{\partial R}\,.\label{eq:filtra}
\end{equation}
Here we have assumed the sound speed varies slowly compared with $\Sigma_{g}$ so that
it can be treated as a constant, which is a good approximation at the gap edges.

In order to proceed, we need to assign a planet-induced gap shape $\Sigma_{g}$, which can be derived by balancing
the planetary torque density and the gradient of the viscous torque (Ward 2009)
\begin{equation}
3\pi\nu R^{2}\Omega\frac{\partial \Sigma_{g}}{\partial R}=\mu^{2}R^{2}\Omega^{2}\Sigma_{g}\frac{R}{x^4}\,,
\end{equation}
where $\mu=M_{p}/M_{*}$, and $x=(R-R_{p})/R_{p}$. Assuming $R$, $\Omega$, and $\nu$ are constant 
(which is a good approximation in the gap region compared with the factor $x^{-4}$),  
this equation can be integrated to obtain the gap profile
\begin{equation}
\Sigma_{g}=\Sigma_{g,0}e^{-|W/x|^{3}}\,, \,\,
W=\left(\frac{\mu^{2}R^{2}\Omega}{9\pi\nu}\right)^{1/3}\,.\label{eq:gapsig}
\end{equation}
where $\Sigma_{g,0}$ is the ambient unperturbed disk surface density \footnote{When a gap is opened, the planetary
torque density also decreases. Assuming the torque density only depends on $\Sigma_{g,0}$, we have overestimated
the torque density so that our gap shape is sharper than a real gap.}.

Plugging Equation \ref{eq:gapsig} into Equation \ref{eq:filtra} and noticing that $T_{s}=\rho_{p}s\pi/(2\Sigma_{g})$, we obtain
\begin{equation}
\frac{d \ln(\Sigma_{d}/\Sigma_{g})}{dR}=\frac{3(W/x)^{4}\rho_{p}s\pi e^{|W/x|^{3}}}{2\Sigma_{g,0}\alpha R_{P} W}\,.
\end{equation}
Note that $T_{s}$ is also a function of $\Sigma_{g}$, suggesting that
dust and gas are more decoupled within the gap.
Integrating this equation with respect to $x$, and assigning $T_{s,0}=\rho_{p}s\pi/(2\Sigma_{g,0})$, we derive
\begin{equation}
-\frac{T_{s,0}}{\alpha}e^{|W/x|^{3}}=\ln {\left(\frac{\Sigma_{d}/\Sigma_{g}}{\Sigma_{d,0}/\Sigma_{g,0}}\right)}=\ln{\left(\frac{\gamma}{\gamma_{0}}\right)}
\end{equation}
where $\gamma$ is the dust to gas mass ratio, $\Sigma_{d}/\Sigma_{g}$, at position $x$ within the gap. 

Then, $x/W$ can be translated back to $\Sigma/\Sigma_{0}$ with Equation \ref{eq:gapsig}, giving
\begin{equation}
-\frac{T_{s,0}}{\alpha}\left(\frac{\Sigma_{g}}{\Sigma_{g,0}}\right)^{-1}=
\ln \left(\frac{\gamma}{\gamma_{0}}\right)\,. \label{eq:gamma}
\end{equation}

The relationship between the dust depletion factor in the gap
 ($\gamma/\gamma_{0}$, the depletion of the dust/gas mass ratio inside the gap compared
 with that outside the gap) and the gaseous gap depth ($\Sigma_{g}/\Sigma_{g,0}$)
is shown in Figure \ref{fig:fig8} with the assumption that $\alpha$=0.01 and $\dot{M}=10^{-8}\msunyr$.  As in \S 4.3.2, if the dust depletion factor is smaller than 0.001,
then we consider that the particles are filtered efficiently. 
From the hydrodynamic simulations, we know the gaseous gap opened by a planet that is a few times more massive than Jupiter is on the order of 
0.01 of  the unperturbed disk surface density (this is clearly shown in the bottom panel of Figure \ref{fig:fig4} where
the gaseous gap opened by a 6 $M_{J}$ planet is 2 orders of magnitude deep.). Thus, Figure \ref{fig:fig8} shows
particles smaller than $100 \mu {\rm m}$ will penetrate through such a planet-induced gap and not be filtered. 

We note that, with a given gap depth and structure, the dust depletion within the gap 
only depends on one dimensionless parameter $T_{s,0}/\alpha$.
Since $T_{s,0}=\rho_{p}s\pi/(2\Sigma_{g,0})$, this parameter is $\propto s/(\Sigma_{g,0}\alpha)$.
Considering that the quantity $\Sigma_{g,0}\alpha$ is proportional to the disk mass accretion rate ($\dot{M}$), 
 $T_{s,0}/\alpha$ depends only on the particle size over the disk accretion rate ($\propto s/\dot{M}$). 
The curve labeled with 10 $\mu {\rm m}$ in our Figure \ref{fig:fig8} assuming $\dot{M}=10^{-8}\msunyr$
can also represent 100 $\mu {\rm m}$ particles in a $\dot{M}=10^{-7}\msunyr$ disk or 1 $\mu {\rm m}$ particles in a $\dot{M}=10^{-9}\msunyr$ disk.
Thus if the disk accretion rate is very low, smaller dust can be filtered.  The reason can be directly seen from Equation \ref{eq:nueta}.
With the same $\Sigma_{g}$ and $T_{s}$, smaller $\alpha$ means less dust diffusion 
so that the critical particle size for which diffusion balances outward drift decreases.
One caution is that the gap density (Equation \ref{eq:gapsig}) diverges to infinity when $x=0$ and in reality the gap density profile truncate at some radius $x$. 
Here we rely on numerical
simulations to determine where the gap truncates and the depth of the gap ($\gamma/\gamma_{0}$). 

Now we consider the second effect: amplified gas radial velocity at the gap edge. The particle drift velocity due to the
pressure gradient (the second term of Equation \ref{eq:driftv3})
 not only needs to be larger than the diffusion velocity as above, but also needs to counter 
the gas radial velocity ($v_{g}$) within the gap. 
Here we compare the dust radial drift velocity derived above with
the gas radial velocity from the gas accretion.
The dust drift velocity (the right side of Equation \ref{eq:nueta}) at the gap outer edge (Equation \ref{eq:gapsig}) is
\begin{equation}
v_{drift}=-\eta V_{K}T_{s}=\frac{3c_{s}^{2}T_{s,0}W^{3}}{R_{p}\Omega x^{4}}e^{|W/x|^{3}}
\end{equation}
Considering the gas flow velocity $v_{g}$ outside the gap gives  $v_{g,0}=-3\nu/(2R)$, and 
the gas velocity is amplified by $v_{g}=v_{g,0}\times\Sigma_{g,0}/\Sigma_{g}$
at the gap edge to maintain a constant accretion rate, we derive
\begin{equation}
\left|\frac{v_{drift}}{v_{g}}\right|=2\frac{T_{s,0}}{\alpha}(W/x)^{4}\frac{1}{W}\,.
\end{equation}
Again we can relate $W/x$ with the gaseous gap depth $\Sigma_{g}/\Sigma_{g,0}$ from 
Equation \ref{eq:gapsig} and obtain
\begin{equation}
\left|\frac{v_{drift}}{v_{g}}\right|=2\frac{T_{s,0}}{\alpha}\left[\ln\left(\frac{\Sigma_{g}}{\Sigma_{g,0}}\right)\right]^{4/3}\frac{1}{W}\,.
\end{equation}
This is plotted in Figure \ref{fig:fig7}. When $|v_{drift}/v_{g}|$ is smaller than 1, 
dust can pass through the gap with the gas due to the amplified gas velocity. 
Thus  with a gap that is two orders of magnitude deep,
10 $\mu {\rm m}$ particles
can pass through the gap with the gas. 
Again the dimensionless parameter $T_{s,0}/\alpha$ is present in this analysis. Here lower $\alpha$ means
lower radial velocity and easier particle filtering. But besides that,
another dimensionless parameter $W$ is also present, and $W$ depends on $\alpha$, $\mu$,
and $R$. At smaller radii, dust is easier to filter since the outward drift velocity is larger.  
For the fixed radii, the dependence of $W$ on $\alpha$ is rather weak.

When 
$|v_{drift}/v_{g}|<1$, more dust particles can penetrate the gap than that estimated
by Equation \ref{eq:gamma} which just considers dust diffusion. Thus,
 when the amplified gas velocity is important, Fig \ref{fig:fig8}
only shows the lower limit of the depletion factor
(marked as the thin curves). On the other hand when 
the the dust diffusion dominates, Fig \ref{fig:fig8}
shows the exact value of the dust depletion factor (marked as the thick curves).

Overall, for a planet (several M$_{J}$) present at 20 AU in a disk with $\alpha$=0.01 and $\dot{M}=10^{-8}\msunyr$, 
the gap outer edge can filter particles larger than 0.1 mm,
but particles smaller than 0.1 mm can pass through the planet-induced gap freely.
If $\dot{M}$ can be lowered to $10^{-9}\msunyr$, this critical size can decrease to $\sim$ 0.01mm.

\section{Discussion}

\subsection{Comparison with previous work}
Paardekooper \& Mellema (2006) have carried out two fluid simulations similar 
to our approach but using a conservative Godunov scheme to study
the response of mm sized dust to the planet. They focus
on the minimum planet mass to open a gap in the dust disk and find a 
0.05 M$_{J}$ planet can open a gap in the mm-sized dust component.
A similar result
was also found using three dimensional SPH simulations by Fouchet et al.(2007). 
This can be partly
understood as the thermal mass ($Gc_{s}^{3}/\Omega$) of a 
dust disk is significantly lower than that
of the gaseous disk since the dust is a pressureless system, so that a gap is easier to be opened in the dust disk. 
 Dust diffusion is ignored in both of these studies. However if the relative density between
 the dust and gas has a large gradient across the gap or spiral wakes, dust diffusion may play an important role
 to smear out any feature in the dust disk.
 Our work focuses
on gap opening by more massive planets to study the effect of the gaseous gap edge on
much smaller grains ($\sim 10 \mu$m) which are essential for near-IR deficit of transitional disk systems.

This dust filtration process by the planet-induced gaseous gap outer edge was first studied by
Rice \etal (2006) with an analytical approach and they concluded micron 
sized particles can be filtered by the planet-induced gap. In detail they suggested that a gap opened 
by a 5 M$_{J}$ planet might be able to filter 1 $\mu$m particles and above. Dust diffusion due to disk turbulence
is ignored in their calculations. Beyond the analytical approach,
 numerical simulations are difficult
to carry out to study dust filtration since small particles have very short stopping time
which limits the numerical time step. In this work we use the SFT approximation
which allows us to study this problem using numerical simulations. 
We have found
that the critical size for dust filtration is larger
 than that estimated by Rice \etal (2006).
In our simulations, the gap outer edge near a 6 M$_{J}$ planet can only filter 100$\mu$m particles
and above. The difference partly comes from a lower accretion rate in their models (the discussion of the critical particle
size on disk accretion rate is in \S 4.4). But
more importantly, the difference is  due to dust diffusion from disk turbulence (Ward 2009).

The larger critical particle size for filtration poses challenges to explain transitional disks with dust filtration alone, as discussed in \S 6. 

\subsection{Steady state, feedback and outer disk}
\begin{figure}
\includegraphics[width=0.4\textwidth]{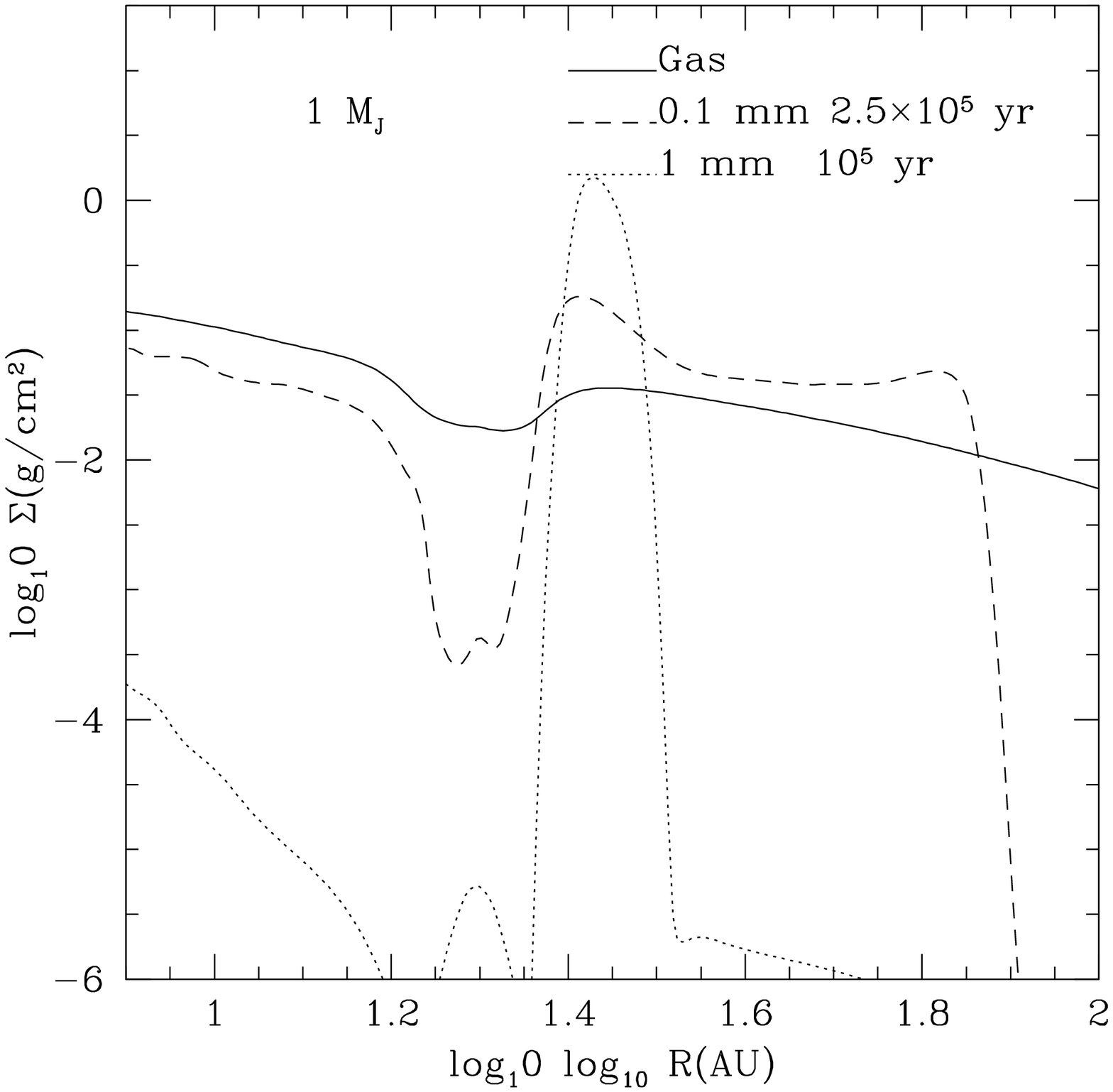} \hfil
\includegraphics[width=0.4\textwidth]{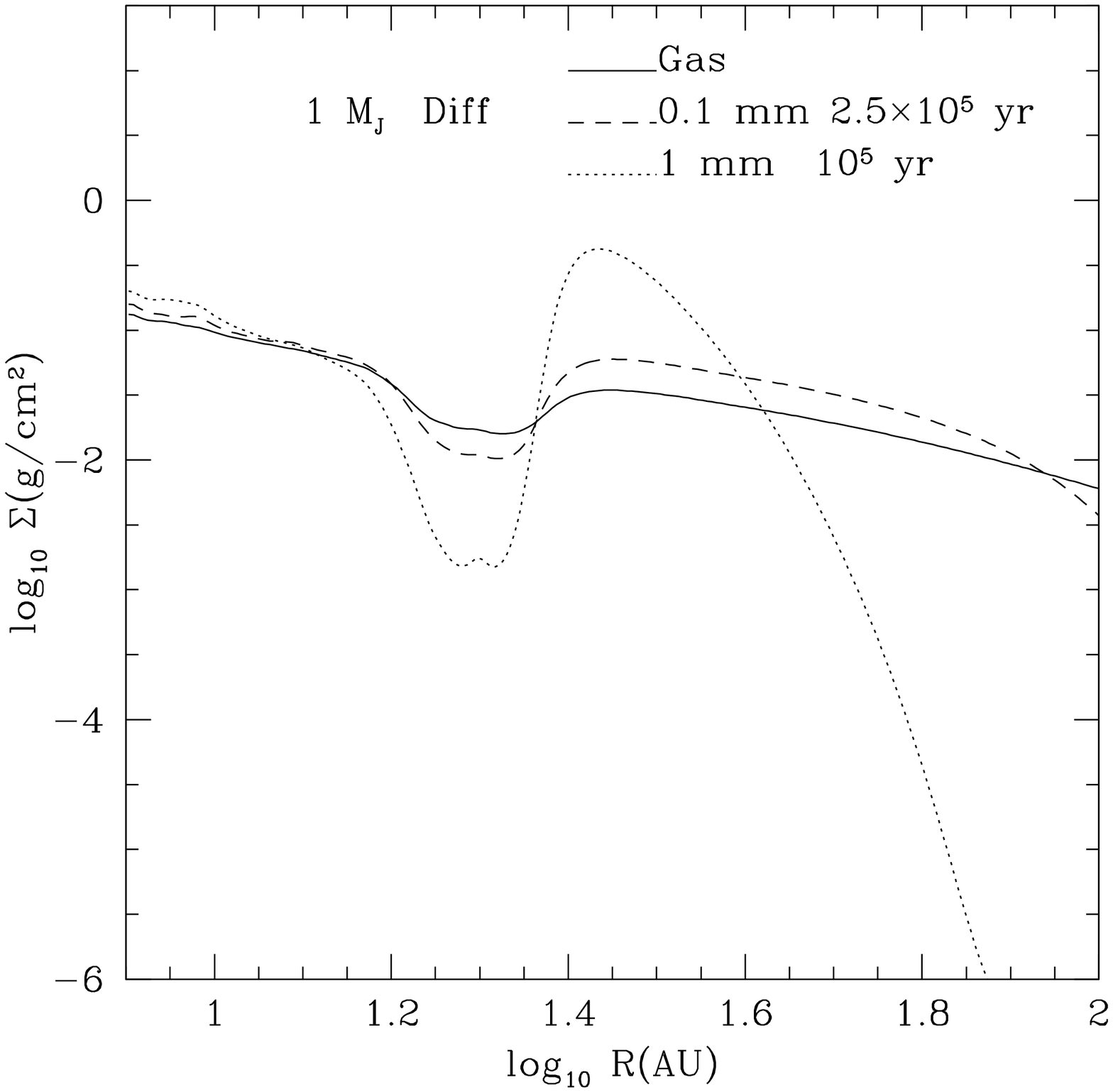} \\
\vskip 0.2cm
\includegraphics[width=0.4\textwidth]{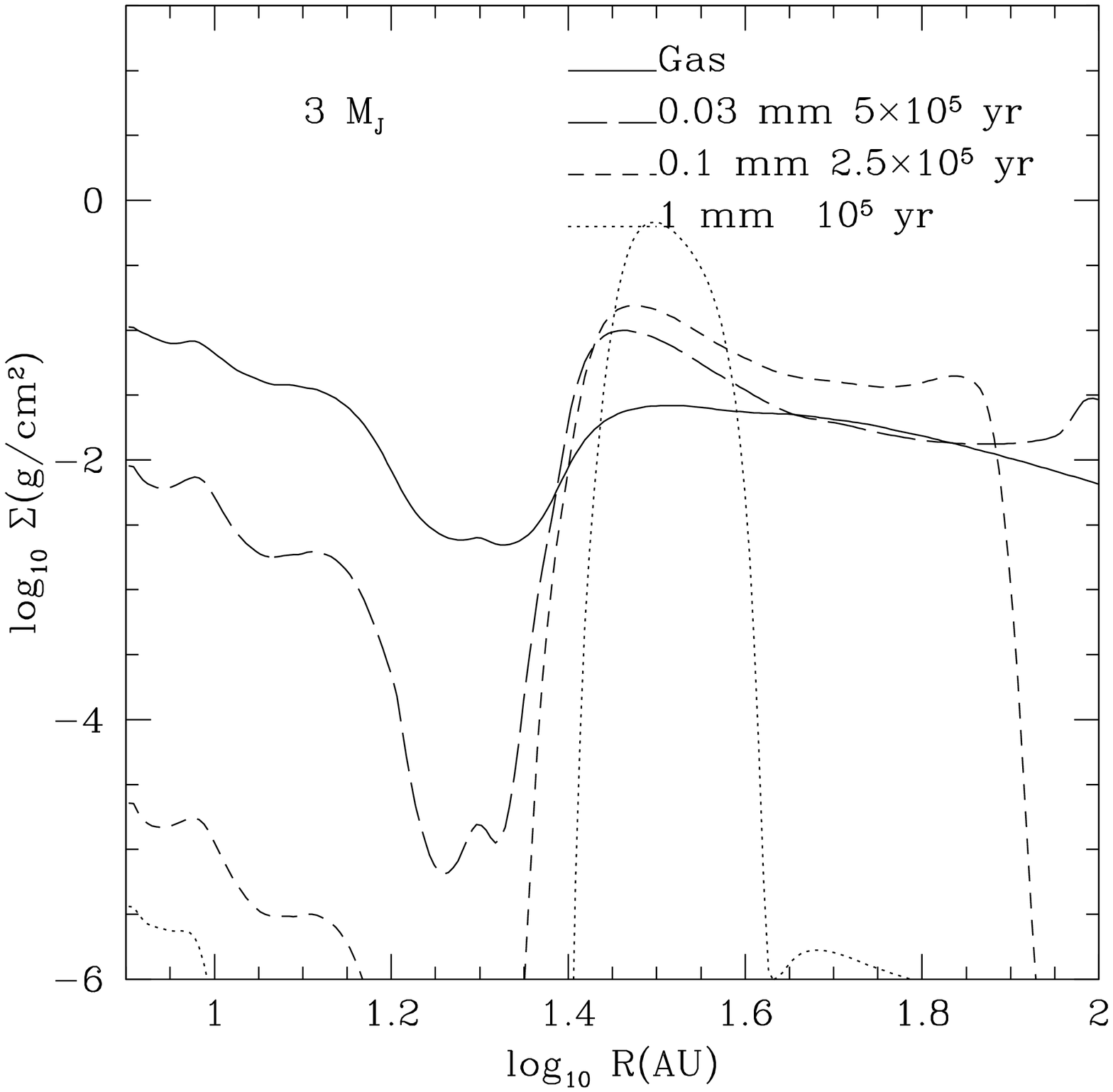} \hfil
\includegraphics[width=0.4\textwidth]{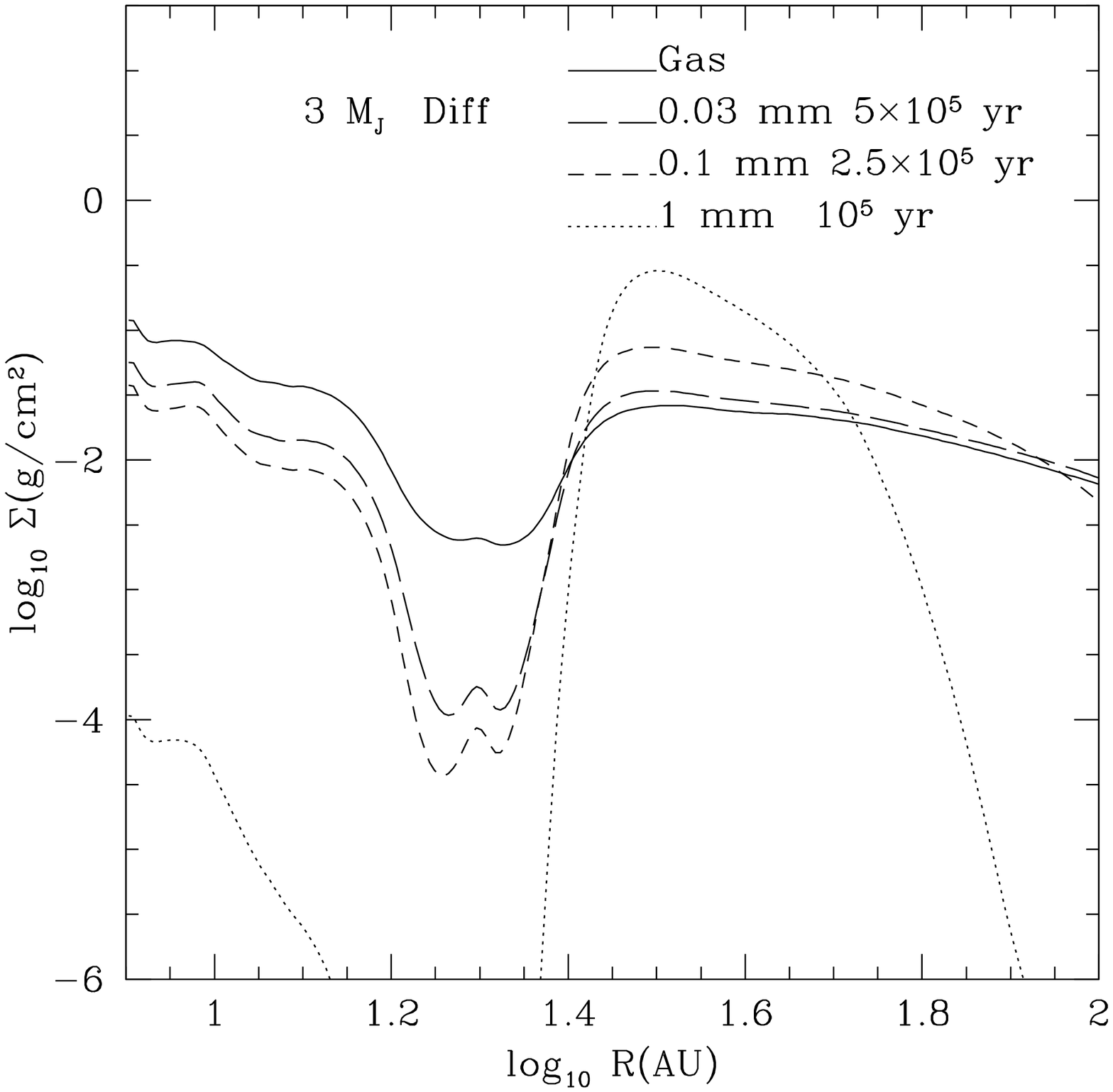} \\
\vskip 0.2cm
\includegraphics[width=0.4\textwidth]{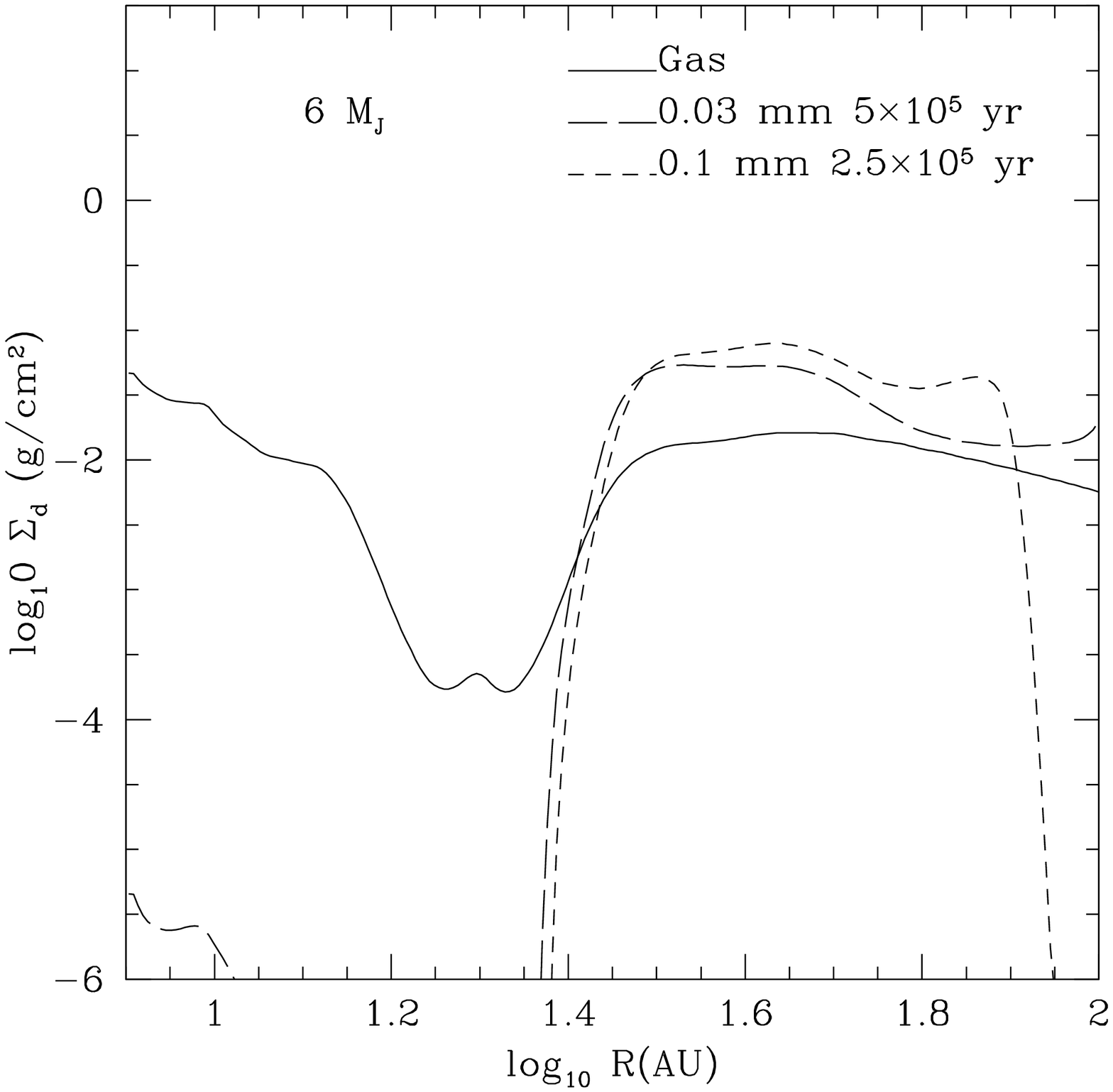} \hfil
\includegraphics[width=0.4\textwidth]{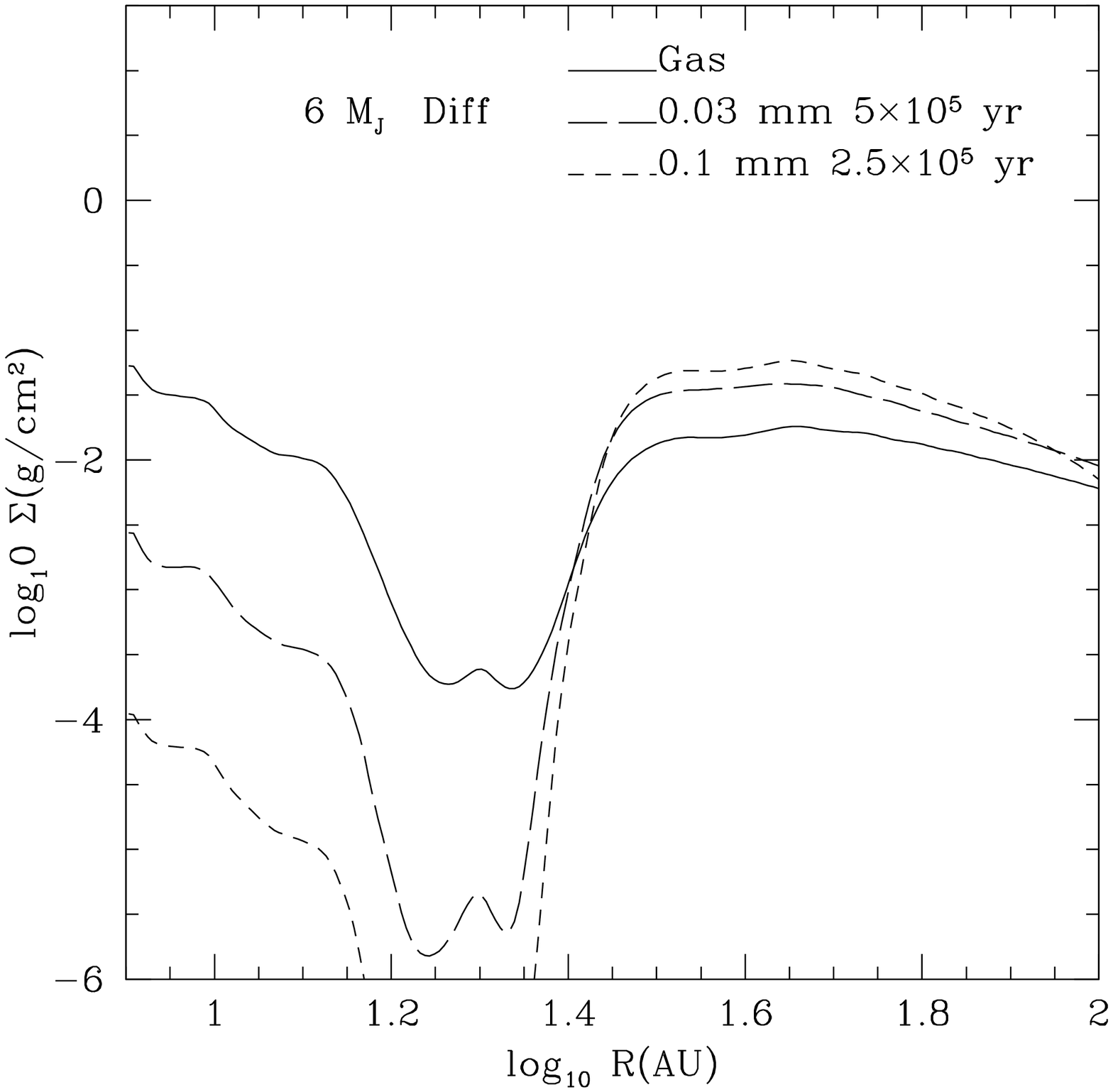} \\
\caption{The azimuthal averaged gas (solid curves) and dust surface densities (
1 mm: dotted curves, 0.1 mm: dashed curves, 0.03 mm: long dashed curves) if the planet is at 20 AU with  different masses
(1 M$_{J}$ upper panels, 3 M$_{J}$ middle panels, and 6 M$_{J}$ lower panels).
 The left panels are without dust diffusion, while the right panels are with dust diffusion considered. 1 mm
 particles are shown at 10$^{4}$ yrs which is longer than its radial drift timescale, while 0.1 mm and 0.01 mm
 particles are shown at 2.5$\times$10$^{5}$ yrs and 5$\times$10$^{5}$yrs since their drift timescales are longer.
 The gas surface densities are divided by 100 to scale with the dust surface densities. Smaller particles can be filtered
 by a gap induced by a more massive planet. However, particles smaller than 0.1 mm cannot be filtered even with a 6 M$_{J}$
 planet if dust diffusion due to disk turbulence is properly included.}
\label{fig:fig4}
\end{figure}

\begin{figure}
\includegraphics[width=0.42\textwidth]{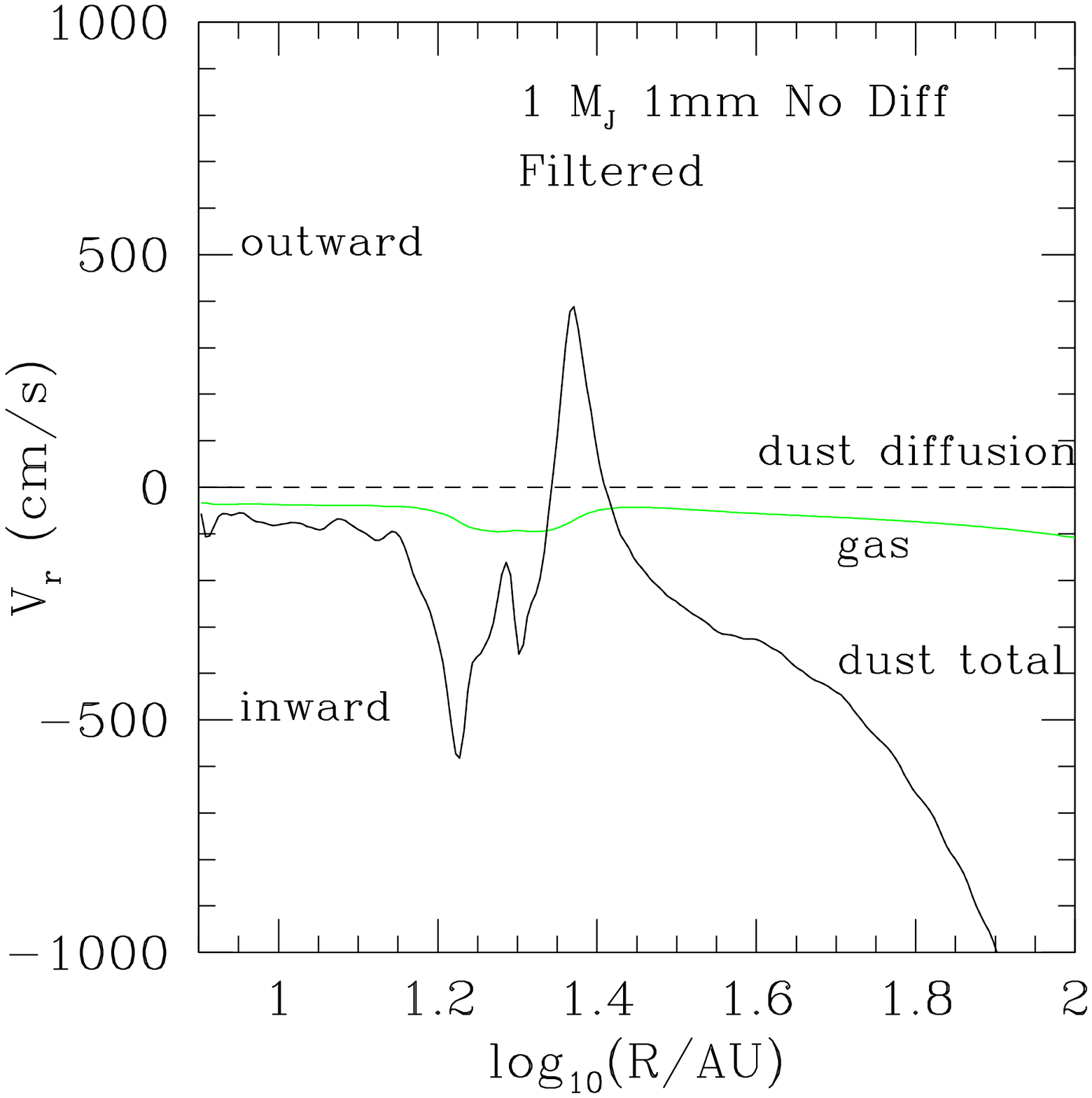} \hfil
\includegraphics[width=0.42\textwidth]{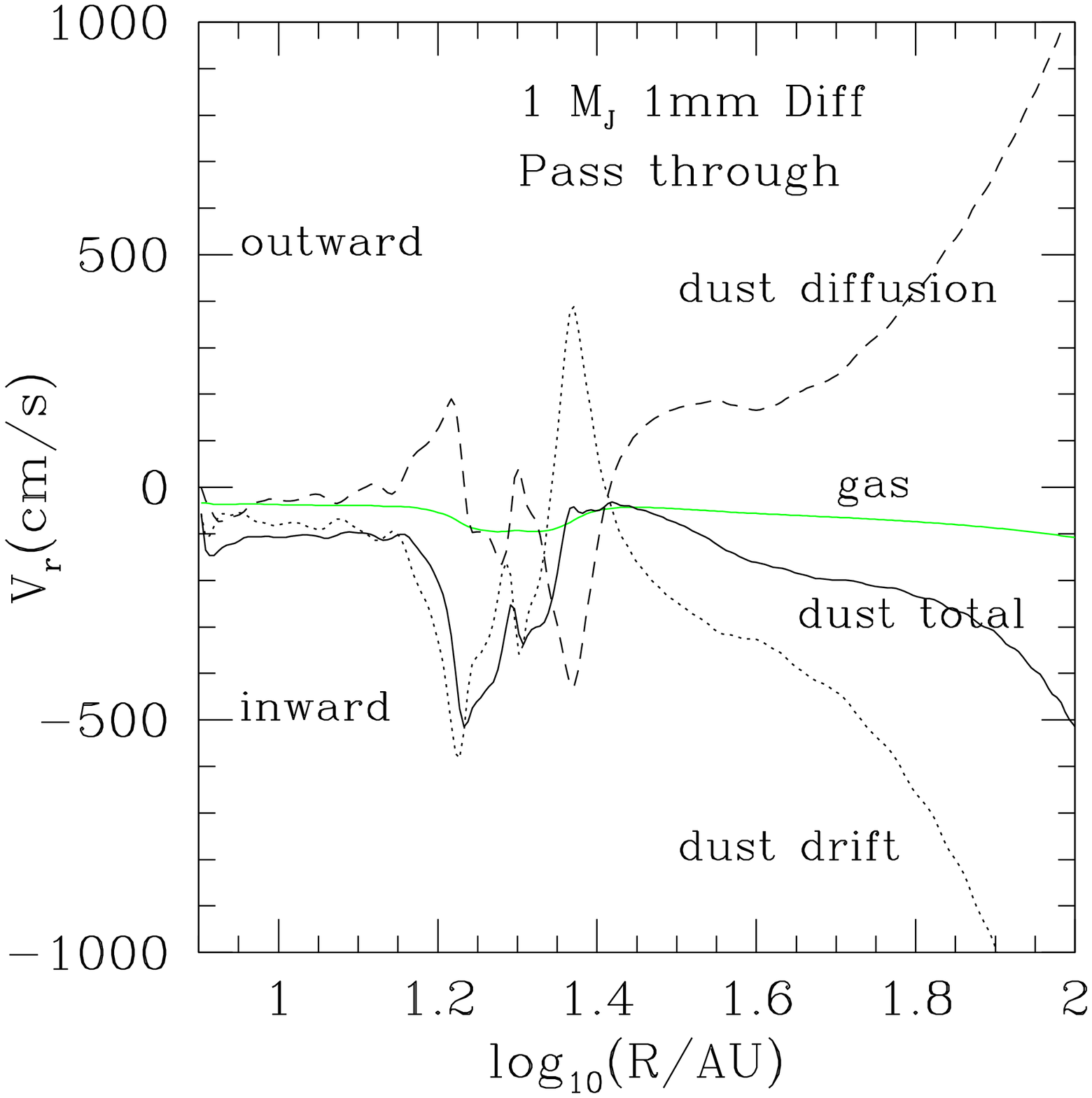} \\
\caption{The 1 mm dust radial drift velocity from 1-D models with (right panel) 
and without (left panel) dust diffusion considered. The green curves are the gas radial
velocity. The solid curves are the total dust radial velocity, among which
the dotted curves are the component due to the pressure gradient and gas velocity, and
the dashed curves are the component due to dust diffusion.
Thus, if 
dust diffusion is ignored, 1mm particles can be trapped by the outer edge of the
gap opened by a 1 M$_{J}$ planet (the dust velocity is positive at the outer gap  edge
in the left panel). But with dust diffusion, 1mm particles can pass through the
gap (the total dust velocity is negative at the gap outer edge in the right panel). }
\label{fig:fig5}
\end{figure}

\begin{figure}
\includegraphics[width=0.42\textwidth]{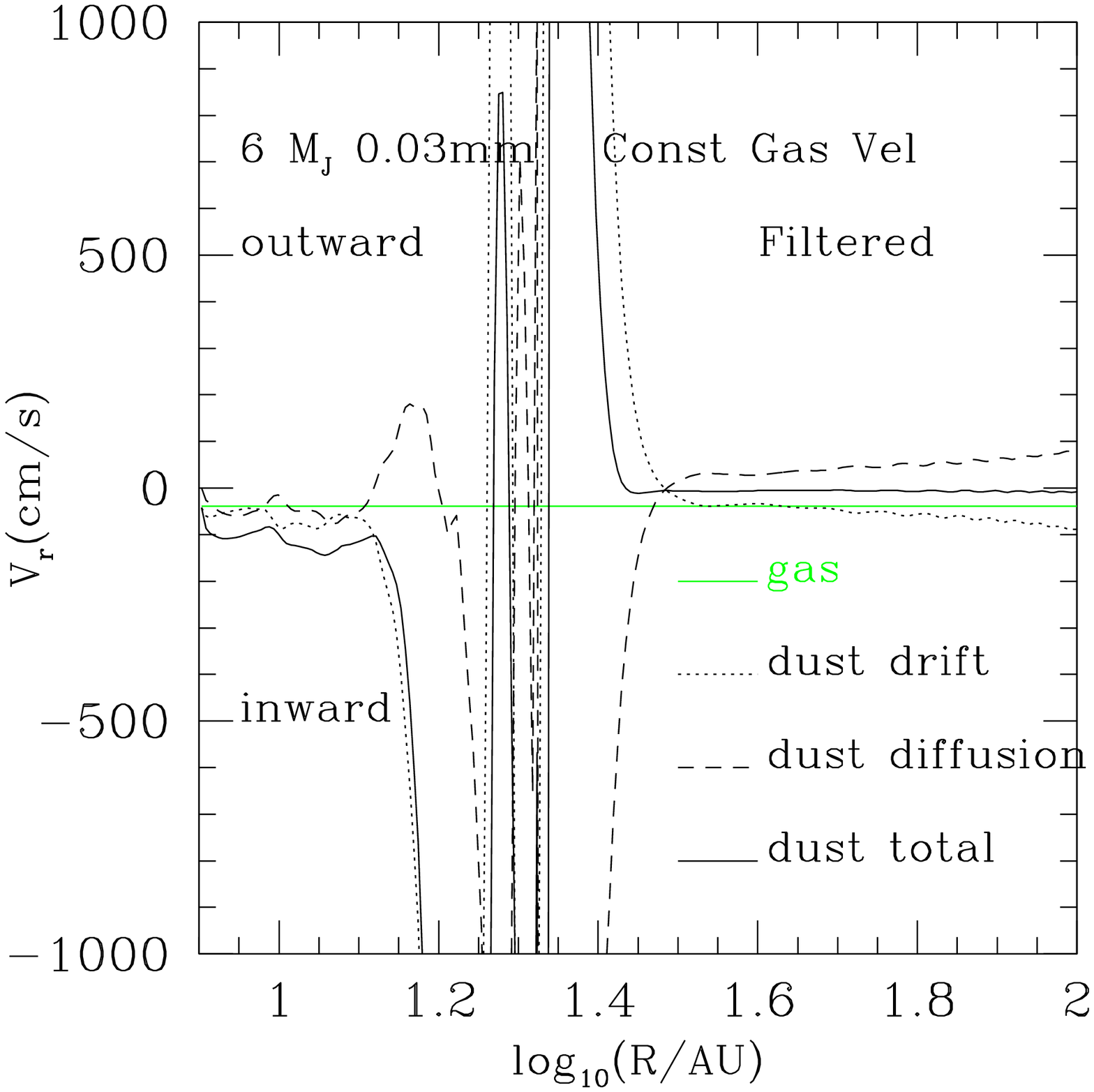} \hfil
\includegraphics[width=0.42\textwidth]{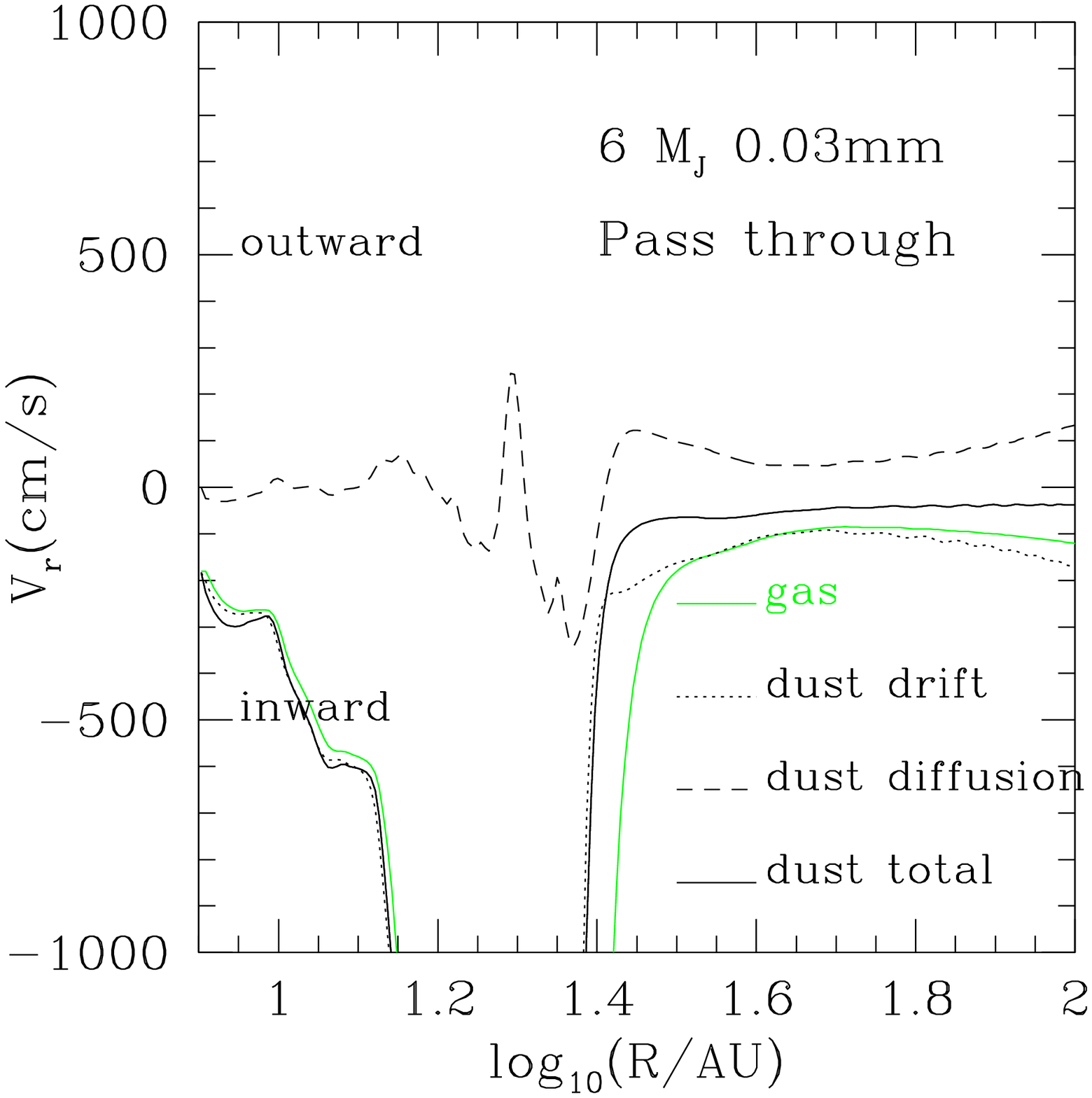} \\
\caption{The 0.03 mm dust radial drift velocities from 1-D models with a 6 M$_{J}$ planet at 20 AU. Dust diffusion 
has been considered in both cases. The simulation
in the left panel artificially assumes the gas velocity is constant across the planet-induced gap, while the simulation
in the right panel correctly assumes the gas mass accretion rate is constant across the gap. 
The green curves are the gas radial
velocity. The solid curves are the dust radial velocity, among which
the dotted curves are the component due to the pressure gradient and gas velocity, and
the dashed curves are the component due to dust diffusion (the significant velocity variation
inside the gap in the left panel is due to the small structures at the corotation region and the incorrect  1-D treatment there, see Appendix). Thus, if the
the amplification of the gas velocity at the gap outer edge is ignored, 0.03mm 
particles can be trapped by the gap edge (the dust velocity, solid curve, is positive at the gap outer edge
in the left panel). But with the amplification of the gas velocity is correctly considered, 0.03 mm particles can pass through the
planet-induced gap (the dust velocity is negative at the gap outer edge in the right panel).}
\label{fig:fig6}
\end{figure}

\begin{figure}
\epsscale{.80} \plotone{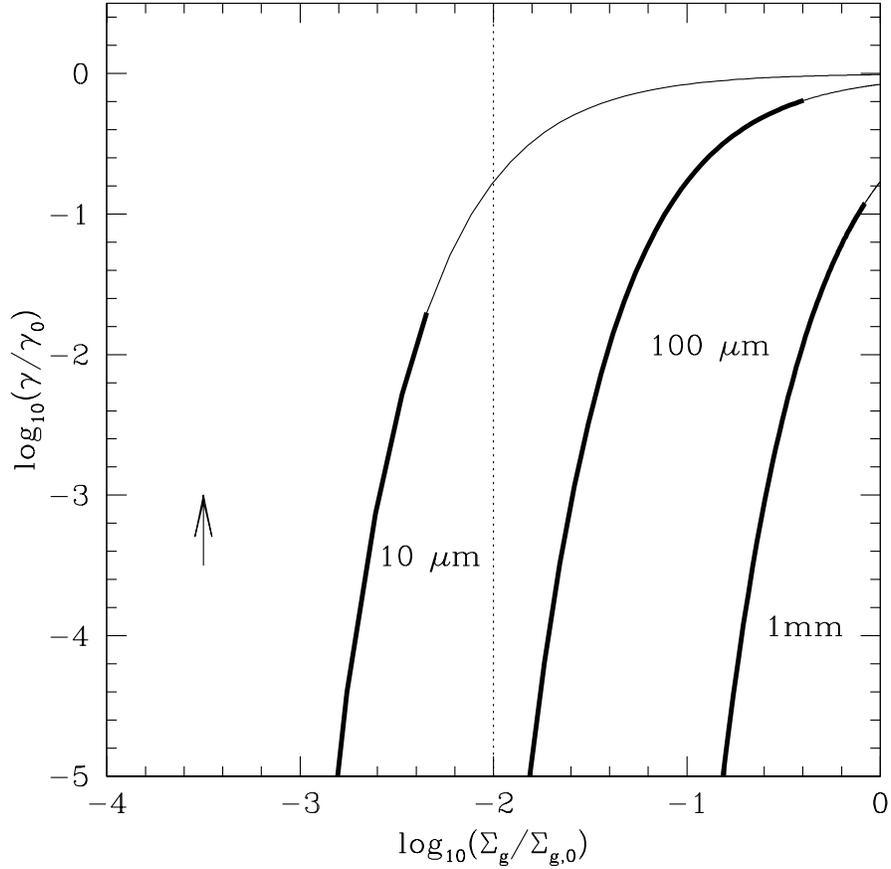} \caption{The dust depletion factor due to dust filtration after dust passes through the gap 
(the depletion of the dust to gas mass ratio with respect to a full disk) with respect to the 
 gap depth for different sized particles. Dust diffusion is considered. The thick 
solid curves are the exact value while the thin curves are just the lower
limit since the gas velocity amplification at the gap edge is important there (Fig. \ref{fig:fig7}). 
The gap depth of 10$^{-2}$ which is opened by a planet with a few Jupiter mass in a $\alpha$=0.01 disk is also labeled.
The arrow indicates the dust depletion factor 1000, which is used by us to determine if dust will 
be filtered. Clearly, particles smaller than 100 $\mu$m cannot be filtered. This plot can be partly scaled to other disk
parameters using the dimensionless parameter $T_{s}/\alpha$. 
} \label{fig:fig8}
\end{figure}

\begin{figure}
\epsscale{.80} \plotone{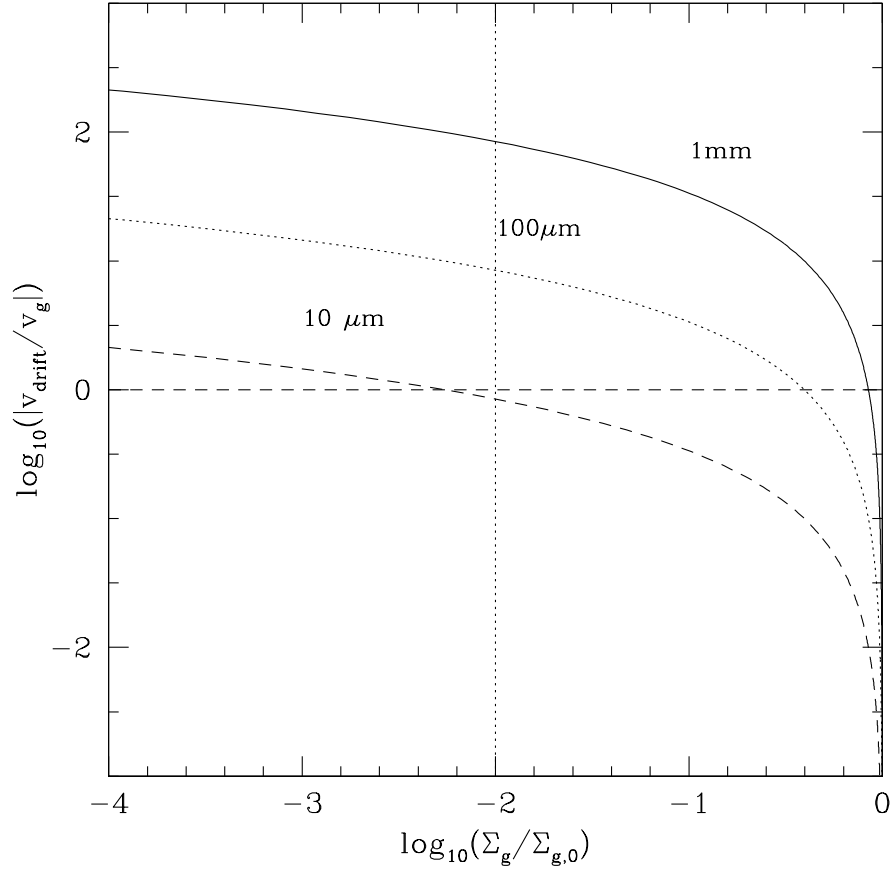} \caption{
The relationship between the ratio of the dust drift velocity over the gas velocity  
and the gap depth. Dust diffusion is ignored in this calculation. If the ratio is larger than 1, particles will be trapped.
Otherwise, particles will be carried inwards through the planet-induced gap by the gas. 
Different sized particles (1 mm, 100$\mu$m, and 10$\mu$m) 
have been considered. Note that the gas velocities are amplified in the gap.   
The gap depth of 10$^{-2}$ which is opened by a planet with a few Jupiter mass in a $\alpha$=0.01 disk is also labeled.
} \label{fig:fig7}
\end{figure}

Without considering dust feedback and dust diffusion, the dust velocity is fully determined by the gas surface density 
(Equation \ref{eq:driftv}). Thus, due to the lack of feedback between the dust concentration
and its velocity, the dust disk cannot achieve a steady state. Dust will continue to pile 
up at the outer edge of the gap (e.g. the upper left panel of Fig. \ref{fig:fig4}), and 
eventually dust feedback on the gas through drag forces becomes important there, 
as pointed out by Ward (2009).

However, with dust diffusion considered, the dust concentration can affect the dust velocity 
(Equation \ref{eq:driftv2}),
and a steady state can be reached on the dust drift timescale. This leads to a smoother and
lower dust surface density (e.g. the upper right panel of Fig. \ref{fig:fig4}), which weakens the 
dust feedback.
Furthermore, with smaller and smaller particles, dust diffusion is much more important than the 
dust drift velocity so that the dust surface density is smoothed even more.
 
The dust disk steady state can be calculated by assuming the product of the dust velocity and density is a constant.
At the outer disk and for 1 mm particles, dust drift and diffusion velocities are far larger than the gas velocity. Thus
we can seek a solution for which the last two terms of Equation \ref{eq:driftv3} balance each other. 
It can be easily derived that a gaseous disk surface density varying as $\Sigma_{g}\propto R^{\beta}$ 
requires the dust surface density to vary as $\Sigma_{d}\propto R^{2\beta-4.5}$ if outward diffusion 
balances inward drift. 
If $\beta$=1, $\Sigma_{d}\propto R^{-2.5}$, which is consistent with our simulations with dust diffusion at the outer disk
(the  left panels of Fig. \ref{fig:fig4}). This property has important implications for submm observations in protoplanetary
disks. It suggests the dust disk will not shrink indefinitely if there is dust diffusion and we can use the dust surface density
structure to imply the gas surface density structure.

\subsection{How much dust can be filtered by a gap in a protoplanetary disk?}
Since our simulations have determined the critical dust size due to gap edge filtration,
we can estimate the fraction of dust mass being filtered by the gap in a protoplanetary disk.

The exact mass fraction of dust larger than some size depends on the dust size distribution function
$ n(s) \propto s^{-\beta}$, where $s$ is the dust size. Dust in the diffuse 
interstellar medium is thought to have a size distribution $\beta=3.5$ 
from 0.005 to 1$\mu$m (Mathis, Rumpl \& Nordsieck 1977). In protoplanetary disks, the size distribution
function can be flatter, possibly due to dust growth, with $\beta=2.5$ (D'Alessio \etal 2001). The total
dust mass fraction for particles smaller than $s_{p}$ is
\begin{equation}
\frac{m(s<s_{p})}{m(total)}=\frac{s_{p}^{4-\beta}-s_{min}^{4-\beta}}{s_{max}^{4-\beta}-s_{min}^{4-\beta}}\,\,\,\, (\rm{if}\,\, \beta\ne4)\,,
\end{equation}
where $s_{max}$ and $s_{min}$ are the maximum and minimum sizes of the dust with distribution 
$n(s) \propto s^{-\beta}$. If both $s_{max}$ and $s_{p}$ are far larger than $s_{min}$, this reduces
to $(s_{p}/s_{max})^{4-\beta}$.

Our simulations above suggest only dust equal or larger than 0.01$-$0.1 mm can be filtered. Thus
if we adopt $s_{max}=1 $mm (D'Alessio \etal 2001), the dust smaller than 0.1 mm only accounts for
10$\%$ of the total dust mass with $\beta=3$ or 1$\%$ of the total dust mass with $\beta=2$, and dust
smaller than 0.01 mm only accounts for 1$\%$ of the total dust mass with $\beta=3$. Thus the dust
mass fraction is decreased significantly when dust crosses the planet-induced gap. With only 1$\%$ dust mass
passing through the gap, the dust to gas ratio within the gap decreases to 10$^{-4}$. However, 
since all the micron-sized dust grains can pass through the gap freely,
the near-IR SED of the disk is hardly affected by the filtration process, which will be 
emphasized in \S 6.

\section{Transitional Disks}
As summarized in the introduction, (pre-)transitional disks have 
wide and deep gaps, and a moderate gas accretion rate onto the star.
Gap opening by planet(s) is an intriguing possibility. To produce these wide and deep
gaps with planets, two main scenarios have been proposed:
gap opening by multiple planets (e.g. Zhu \etal 2011, Dodson-Robinson 
\& Salyk 2011) and dust filtration (Rice \etal 2006).

Since both gap opening by multiple planets (Zhu \etal 2011) and dust filtration (this work) 
hardly affect the micron sized  particle distribution within AU scales, the near-IR SEDs of these disks
should look similar to those of classical T-Tauri disks. Thus both scenarios can explain pre-transitional disk near-IR SEDs,
especially for those having the same SED as classical T-Tauri disks (Andrews \etal 2011).
However, both of the scenarios have difficulties in reproducing transitional disk SEDs 
which have strong near-IR deficits.

In the following
we will use a Monte-Carlo radiative transfer model to demonstrate this difficulty with both
1) gap opening by multiple planets, 2) dust filtration, and provide one possible solution
with 3) filtration+grain growth to reproduce the transitional disk GM Aur's SED.

\subsection{Monte-Carlo radiative transfer set-up}
The Monte Carlo radiative transfer code was developed by Whitney et al.
(2003a,b), Robitaille et al. (2006), and Whitney et al. (2012), while for
the disk structure see Whitney et al. (2003b) for references.
This entire disk is composed of two dust components: a thick disk with small
(i.e., ISM-like $\mu$m-sized) grains, and a thin disk with large (mm-sized)
grains. 
We use the standard ISM dust model for the small-dust (Kim et
al. 1994), and use Model 3 in Wood et al. (2002) for the large-dust 
($\beta$=3 with the maximum particle size 1 mm). 
Both disk components are isothermal in the vertical
direction, and their respective scale heights $h_{\rm thin}$ and $h_{\rm
thick}$ obey a simple power law $h\propto R^\Psi$, with $h_{\rm thin}$
fixed to be $0.2\times h_{\rm thick}$. The disk extends from the
sublimation radius (self-consistently determined by the dust sublimation
temperature) to 200AU. 

To reproduce a classical T Tauri disk SED for comparison, 
we set-up a full disk model, where
the gas surface density
profile is 
$\Sigma_{R}=\Sigma_0\frac{R_c}{R}e^{-R/R_c}$
and $R_c$ is the scaling length fixed to be 100 AU. This is the same as our
hydrodynamic simulations. Dust mass is 1$\%$ of the disk mass.
Among all the dust, 1$\%$ is in the small-dust, while the rest is in the large-dust, which is equivalent
to saying that the small-dust depletion and settling factor is 0.01. This choice is based on both
our argument in \S 5.3 (if the dust distribution function is $n(s) \propto s^{-3}$ from submicron
sizes to 1 mm, 99$\%$ of the dust will be in the grains larger than 0.01 mm), and 
observational constraints (Furlan \etal 2006).
Note that this
implies dust has already grown to mm sizes in the outer disk and the small-dust
is only 1$\%$ of the total dust mass.  
The full disk model has a total mass of 0.1 M$_{\odot}$, and a scale height profile
with $H/R = 0.075$ at 100 AU and $\Psi$=1.2. The accretion rate is assumed to
be 10$^{-8}\msunyr$. These are nominal values for a classical T Tauri disk, and they
are consistent with our hydrodynamical simulations. Since
we are trying to fit the SED of GM Aur, the central
source is assumed to be a 5730 K pre-main star with radius 1.5 R${_\odot}$
and mass 1.2 M$_{\odot}$ (Calvet \etal 2005) and the inclination angle of this system is 55 degrees.

\begin{figure}
\epsscale{.80} \plotone{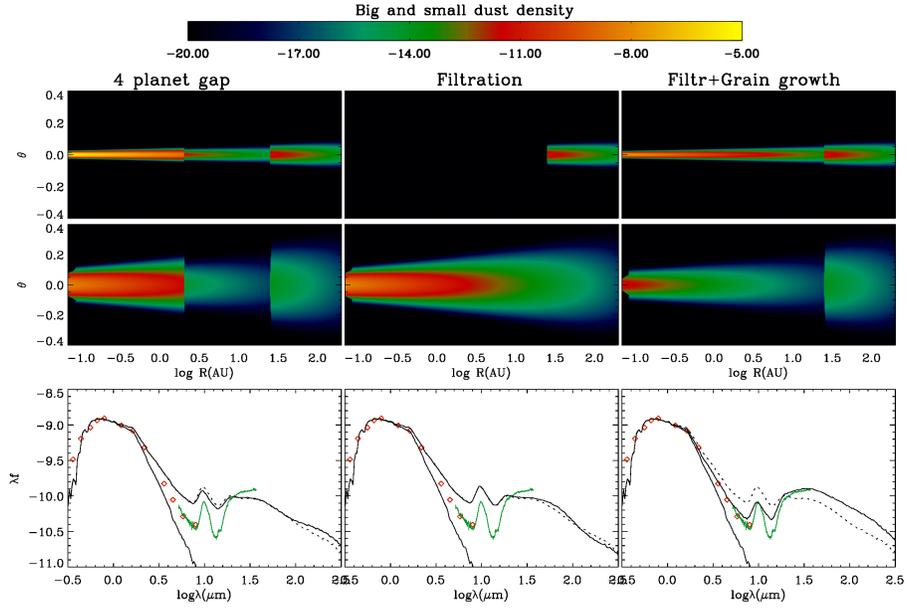} \caption{ Various scenarios to explain transitional disks 
GM Aur.
1) a wide gap opened by multiple planets (left panels); 2) a deep gap opened by one planet
which can filter large dust particles (middle panels); and 3)  After big particles are filtered by the gap,
small particles can grow (right panels). The upper panels show the dust density distribution for large particles ($\gtrsim$10 $\mu$m) in the disk (big
particles are totally filtered in the second scenario and some new big particles are generated in the third scenario), 
while the middle panels show the dust density distribution for small particles ($\lesssim$10 $\mu$m) in the disk (small-dust
is continuous in the second scenario). The lower panels show the SED from these models. The dotted curve
is the SED from a full disk.
Photometric (red, open symbols) and IRS data (green) for GM Aur are from Espaillat et al. 2011; refer to that work for more details. 
As clearly shown, either gap opening by multiple planets or dust filtration (left and right panels)
has little effect on the SED compared with a full disk (solid curves overlap with the dotted curves). This is because 
small dust particles in the inner disk ($<$1 AU, which produce most of the optical and IR flux) 
 still have similar abundance as the full disk model.
But dust filtration plus dust growth can explain
transitional disk SED, since small dust particles in the inner disk ($<$1 AU) are depleted significantly in this scenario.
} \label{fig:fig9}
\end{figure}

The full disk's SED is shown as the dotted curves in the bottom panels of Fig. \ref{fig:fig9}. 
It produces a strong near-IR flux, similar to a classical T Tauri disk SED. In the 
following we will modify this full disk model based on results from hydrodynamic simulations 
trying to reproduce transitional disk GM Aur's SED (the red curves in Fig. \ref{fig:fig9}).

\subsection{Gap Opening by Multiple Planets?}
To simulate gap opening by multiple planets, 
 we cut a wide gap in the disk for both small and
large dust (the left panel of Figure \ref{fig:fig9}). In Zhu \etal (2011),
with the same disk structure,  using hydrodynamic simulations, 
we found that four giant planets can open a gap from 2 to 20 AU
with the gap depth 1/1000 th of the unperturbed disk surface density. Thus
here we cut the gap from 2 to 25 AU and both small and large dust surface
density is decreased by a factor of 1000 compared with the unperturbed disk.
The gap is clearly seen in the dust surface density contours in the left panels of Figure \ref{fig:fig9}.
However, the near-IR SED from the modeled disk with 4 planets produces an 
SED more similar to that of classical T Tauri disks, rather than transitional disks (the left bottom panel).
This is because the inner disk within 2 AU
is the same as a full disk.

This similar SED as that of a full disk is not surprising, since
previous work has already suggested the near-IR deficit of transitional disk requires 
small-dust in the inner disk, on scales less than 1 AU,
to be depleted by many orders of magnitude (Espaillat \etal 2010, Zhu \etal 2011). 
To be more specific, considering those
 transitional disks having accretion rates $\sim 10^{-8}\msunyr$ onto the star, and
using
\begin{equation}
\Sigma_{g}=\frac{\dot{M}}{3\pi\nu}\,,
\end{equation} 
we know $\Sigma_{g}$ is $10^{2}\, {\rm g}/{\rm cm}^{2}$ at 0.1 AU for $\alpha=0.01$. Considering that
the nominal opacity of ISM dust at 10 $\mu {\rm m}$ is 10 ${\rm cm}^{2}/{\rm g}$, the optical depth at 0.1 AU is
10$^{3}$. But SED modeling of transitional disks (e.g. GM Aur) require the disk to be optically thin
at 0.1 AU (Calvet \etal 2005); thus the micron-sized dust, which contributes most to the near-IR flux, needs to 
be depleted at least by three orders of magnitude.  However, gap opening by multiple planets from 2-20 AU
has little effect on the dust distribution within AU scales and thus it won't prevent the near-IR SED 
from being similar to a full disk model.

\subsection{Dust Filtration by the Gap?}
To simulate the effect of dust filtration, large-dust is absent within 25 AU compared with the full disk model (the middle panel of Figure \ref{fig:fig9}).
 All large-dust is assumed to be filtered. However the small-dust can pass the planet-induced gap
 and has the same distribution as a full disk (middle panels of Figure \ref{fig:fig9})\footnote{We
 neglect the narrow gap opened by the planet in this calculation as it will
 not significantly affect the SED.}. Again, although
 the mid-IR
 SED changes a little bit due to the depletion of the large-dust,
 the optical to near-IR SED looks similar to the full disk (the middle bottom panel) since small-dust dominates the optical to near-IR opacity.  

This demonstrates that although dust filtration is efficient at reducing the total dust mass, 
since most of the dust mass resides 
in large dust particles (\S 5.3), it has little effect on the near-IR SED due to the fact that near-IR SED is determined
by the micron-sized particles which can not be filtered by the gap. Although Dodson-Robinson 
\& Salyk (2011) speculated that dust can be trapped in the spiral wakes, trapping in spiral wakes is only a second order effect
compared with the gap filtration, since the density wakes have far smoother density profiles than the gap edge. 
If the gap edge fails to trap micron sized particles, 
the density wakes won't be able to trap these particles. 
Our two fluid simulations (Figure \ref{fig:fig2} and \ref{fig:fig3}) also do not show dust pile up in the density wakes.

However, under
some extreme circumstances (e.g. $\dot{M}\le10^{-9}\msunyr$, the presence of a 10 M$_{J}$ planet), the critical filtration size may decrease
to 
10$\mu \rm{m}$. With a flat dust size distribution (e.g. $\beta$=2 in \S 5.3), dust smaller than 10 $\mu \rm{m}$
could have a mass less than $10^{-4}$ of the total dust mass. Thus, after filtration, small-dust could have a depletion factor
equal to $10^{-4}$, which is close to the small-dust depletion factor ($10^{-5}$) required to explain the near-IR deficit of GM Aur
\footnote{If the gas surface density is very low (e.g. 2 g$\cm^{3}$) the dust depletion factor can be $10^{-3}$ for the disk to be optically thin (Salyk et al. 2007). 
In this case, a large disk viscosity parameter $\alpha$ is required to
explain the observed gas accretion rates.}.

\begin{figure}
\epsscale{.80} \plotone{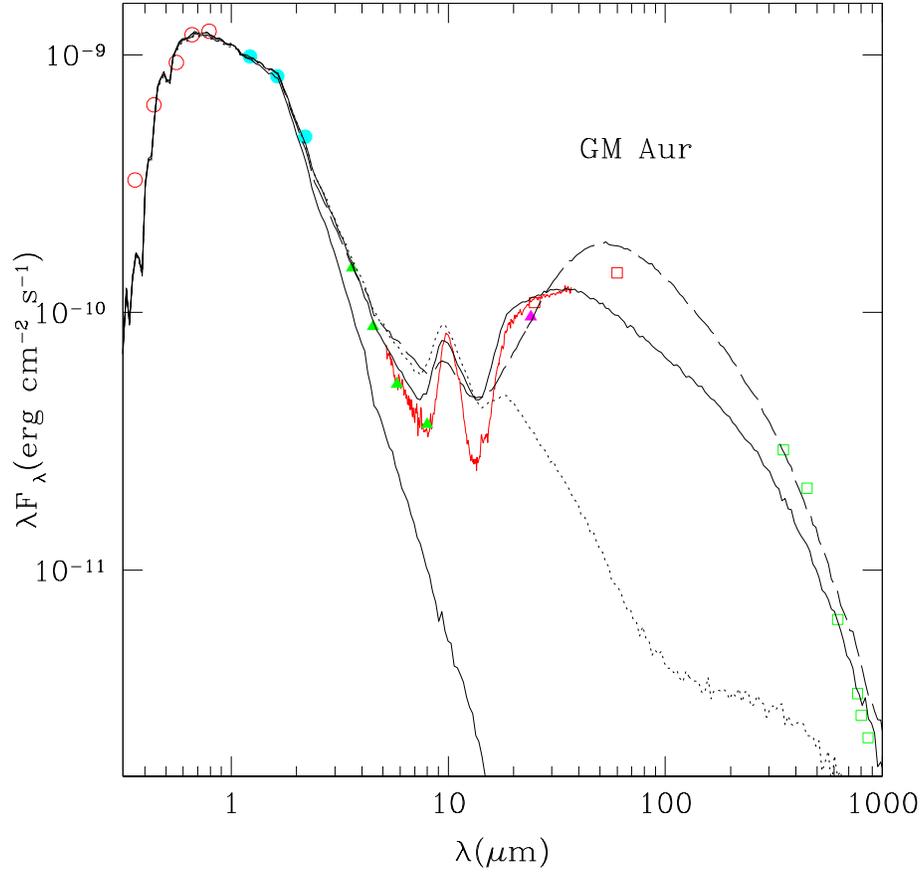} \caption{ The photometry and spectrum for GM Aur (colored dots and curves) and three model SEDs. The solid black curve is from our best fit model as in the right panels of Fig. \ref{fig:fig9} considering both dust filtration and dust growth. The dotted curve is from the dust filtration only model (similar to the middle panels of Fig. \ref{fig:fig9}) but with small/large dust ratio 10$^{-5}$ at the outer disk. This model can reproduce the near-IR flux since, after big particles are filtered by the planet-induced gap, it basically has the same amount of small
dust in the inner disk as our best fit model. However due to the lack of small-dust at the outer disk, the outer disk
is quite optically thin with big particles settle to the midplane (20$\%$ thickness) and produces very little mid-IR flux. The dashed curve is similar to the dotted curve, but the big-dust at the outer disk is not allowed to settle and have the same thickness as the gaseous disk, which is unlikely. 
} \label{fig:fig10}
\end{figure}

In Fig. \ref{fig:fig10}, we have calculated 
two cases with a small($<10\mu$m)/large($>10\mu$m) dust mass ratio $\sim$ 10$^{-5}$ at the outer disk and only small-dust
existing in the inner disk. As expected, in this case, the near-IR deficit can be reproduced due 
to the tiny amount of small-dust in the disk. However, this tiny amount of small dust makes not only  
the inner disk optically thin but also the outer disk's atmosphere optically thin
 (assuming large dust grains have settled to the midplane forming a dust layer whose 
thickness is only 20$\%$ of the gas disk). The mid-IR flux is very weak too and unable 
to reproduce the IRS observations (dotted curve in Fig. \ref{fig:fig10}). In order 
to reproduce the mid-IR flux, the large dust grains have to remain suspended in the 
atmosphere of the disk and not settle (maintaining the same thickness as the gaseous disk) to intercept the stellar radiation, which is  
shown by the dashed curve in Fig. \ref{fig:fig10}). 

The dashed curve in Fig. \ref{fig:fig10} suggests that it is not impossible for dust filtration alone 
to explain GM Aur. But several conditions have to be met: 
1) dust grows significantly in the outer disk (flat dust size distribution, and the 
small-dust ($<$10$\mu$m) to big-dust mass ratio is 10$^{-5}$);
2) a very massive planet forms in a disk with a low accretion rate;
3) large grains in the outer disk cannot settle. These conditions are not easy to satisfy since
a normal T Tauri disk has a small-dust depletion 
factor of 0.1-0.001, and large grains are settled to the midplane. Furthermore GM Aur 
has a disk mass accretion rate $10^{-8}\msunyr$.

\subsection{Filtration+Grain Growth}
As implied above, to reproduce the near-IR deficit of transitional disk SEDs it is essential to reduce the small-dust 
($\le$0.01 mm) abundance in the 
inner disk within AU scales. One solution is considering the growth of small dust particles after large 
particles are filtered by the planet-induced gap. 
Dust can grow quite rapidly in the inner disk. It may only take 10$^{3}$ yrs for dust particles
to grow from sub-micron sizes to 1000 $\mu$m at 1 AU (e.g. models S3 and S4 in Dullemond \& Dominik 2005). One way to stop the rapid dust growth
is by collisional fragmentation (Dullemond \& Dominik 2005, Dominik \& Dullemond 2008), in which case
large particles are shattered to replenish small dust grains. Thus, the growth and fragmentation maintains a quasi-stationary dust size distribution function.   
In disks with gaps, if all big grains ($>10-100\mu$m) are filtered acrossing the gap
(accounting for 99$\%$ of the total dust mass if we assume the critical filtration size is 
0.1 mm with $\beta=2$ or the critical size is 0.01 mm with $\beta=3$, as estimated in \S 5.3 ), 
the remaining small grains (accounting for 1$\%$ total dust mass) 
that manage to pass through to the inner disk
can grow quite efficiently without replenishment from fragmentation of large particles. Although the dust growth time is 
100 times longer than the timescale for a non-filtered disk
(the dust growth time is inversely proportional to the dust abundance), it is still modestly
 shorter than the  at 20 AU with $\alpha$=0.01.  Eventually a new balance is made,
and the quasi-stationary dust size distribution is established again. At this time, due to grain growth, the small-dust is only 
$1\%$ of the abundance that was present after it had passed through the planet's orbit location.  In other words, small-dust 
is depleted indirectly due to dust growth and the net depletion factor
for small particles in the inner disk is 10$^{-4}$($1\%$ after filtration $\times 1\%$ mass fraction in the new size distribution).
Such a scenario of ``double depletion'' may explain transitional disks, but requires further study to place it on a firmer
foundation.

To simulate this scenario we assume small-dust grows in the inner disk and reestablish the
dust size distribution after large particles are filtered by the planet-induced gap at 25 AU. 
This is illustrated in the right panels of Fig. \ref{fig:fig9}. In the inner disk we assume
the new dust size distribution has a small to large dust mass ratio 1:999 \footnote{Ratio 1:99 gives similar results but the near-IR flux
is still a little bit higher than observations}. Since the total dust mass 
is decreased to
1$\%$ after dust filtration and this new distribution further reduces the small-dust by a factor of 1000, the small-dust is depleted by a factor of 10$^{5}$, and the mass ratio of small-dust to
gas is $10^{-7}$. This leads to an optically thin inner disk.
The disk's SED fits the GM Aur SED quite well in the bottom right panel.
 
 Note that in this scenario small particle growth is a gradual process.
 Although we deplete the dust in the inner disk uniformly for computational convenience (upper two panels in the
 right column of Fig. \ref{fig:fig9}, or in other words we assume small particles grow instantaneously after they cross the gap); in reality
 the abundance for small particles may change gradually and join the outer disk abundance smoothly.

The readers may notice 10 $\mu$m silicate feature is not fitted well by our simple models. But we want to point out that 
the strength of the
silicate emission depends not only upon the total amount of dust inside the gap, but upon dust size distribution, especially for small
dust ($<$10$\mu$m).  The dust size distribution sensitively depends on growth, fragmentation, and
differential drift of particles of differing size.  Predicting the strength of the silicate emission feature will require
addressing the very complex processes involved in grain evolution in concert with 
calculations of drift similar to those in the present paper. For example, when small dust particles grow to big particles, big
particles will quickly drift to the central star, and if the drift timescale is much shorter than the dust growth timescale,
only small particles will exist in the inner disk. A careful dust evolutionary model combined with dust filtration is important
to test this scenario. Nevertheless, small-dust needs ($<$10$\mu$m) to be depleted more than that dust filtration predicts to explain near-IR deficit
of transitional disks.
 
In this work
we only put one planet in the disk to study dust filtration, because the dust filtration process does not sensitively depends on the number 
of the planets in the disk since most dust will be filtered by the outermost planet gap or the deepest gap. Thus most our
results can also be applied to the gap opened by multiple planets in the disk, which is complimentary to Zhu \etal (2011).
We note, however, that a multiple planet system may need to be invoked to dynamically clear planetesimals from the inner disk region
as the presence of such bodies is likely to provide a source for small dust grains through their mutual collisions.

\subsection{Observational implications for ALMA}
Dust filtration has other observational implications besides SEDs.
Dust filtration by the planet-induced gap differentiates various particles. Thus if we observe
transitional disks at various wavelengths the gap/cavity should be more distinctive at longer wavelengths
(Fouchet et 
al. 2010 and our Figs. \ref{fig:fig1} to \ref{fig:fig4}). 

Dust growth as argued above won't happen instantaneously as the flow passes through the 
planet-induced gap. Thus at shorter wavelengths, the cavity which is found by submm observations 
is less apparent or even disappears (e.g. Fig. 1). Andrews \etal (2011)
notice that there are (pre-)transitional disks with classical T-Tauri disk SEDs but that
show gaps in submm interferometry.  
More directly, recent Subaru observations have found a lot of (pre-)transitional disks 
that have submm cavities (Andrews \etal 2011) do not show cavities in near-IR scattered light images 
(Dong \etal 2012). Both of these findings seem to agree with the dust filtration scenario.

\begin{figure}
\epsscale{.80} \plotone{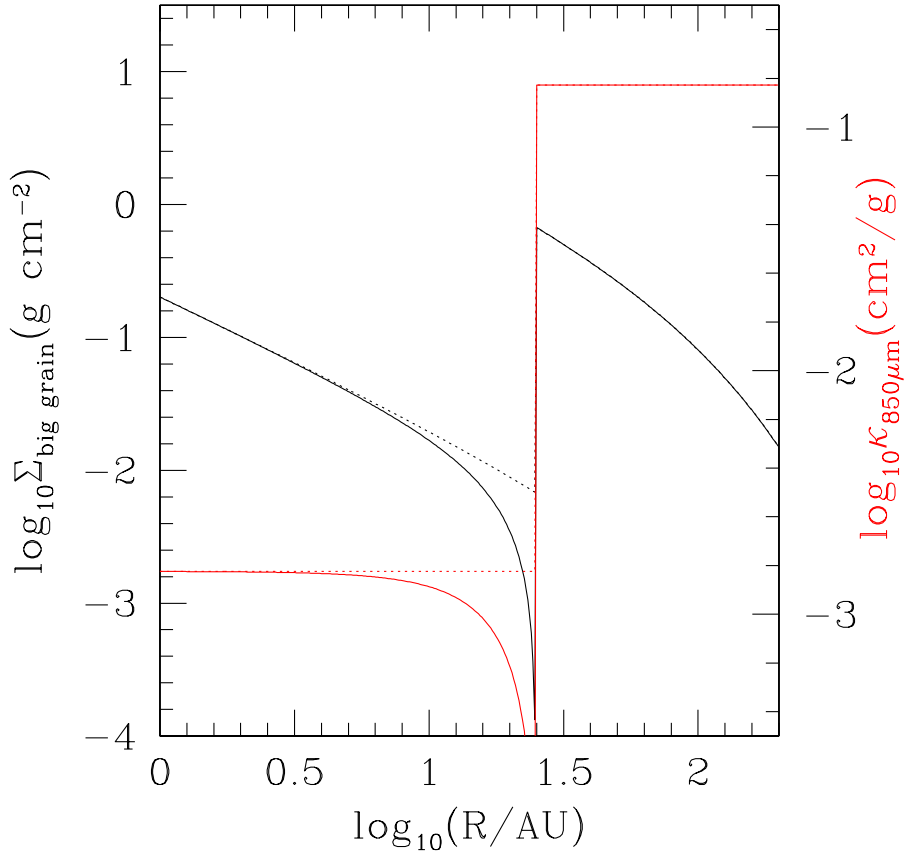} \caption{ The big grain ($\gtrsim$10 $\mu$m) surface density in our simple model of filtration+grain growth scenario (the right panels of Figure \ref{fig:fig9}), and
the 850 $\mu$m opacity (the red curves). The opacity is calculated by dividing the total optical depth at 850 $\mu$m over the gas surface density assuming the gas surface density is the
same as that in the full disk model unaffected by the planet. The decrease at 20 AU for both quantities are due to gap filtering big particles.
The dotted curves are from the model presented in the  right panels of Figure \ref{fig:fig9}, where grains are assumed to grow instantaneously after they pass
through the gap. This is for numerical convenience. The solid curves may be more close to a real protoplanetary disk, where grain growth is a gradual process. However, in this calculation we ignore the radial drift of particles. 
If the radial drift of the big particles is considered, the big-dust surface density (or the   850 $\mu$m opacity) can be either flatter or steeper depending
on both the radial drift timescale and the grain growth timescale.  Although the grain growth of the inner disk is uncertain, 
the sudden decrease of the submm opacity 
at the gap edge and slowly change inwards is a signature for dust filtration, which can be tested by ALMA.
} \label{fig:fig12}
\end{figure}

However, after the big grains are filtered, whether and how much small grains can grow is a difficult issue; 
it requires a dust evolutionary model combined with dust dynamics. The complexity of the problem is illustrated by, for example,
Birnstiel \etal (2011) and references therein. In this work, based on transitional disk SEDs, we 
suggest that small-dust needs to grow in the inner disk.
Figure \ref{fig:fig12} shows both big-dust (larger than the critical filtration size) surface density
and the 850 $\mu$m opacity in the filtration+grain growth scenario\footnote{To calculate 
the opacity, we assume the gas surface density is 
unaffected by the presence of the planet. The gaseous disk is the same as a constant 
$\dot{M}$ accretion disk.}. 
Both the large-grain surface density and the submm opacity change at the gap edge
due to dust filtration. The submm opacity decreases by a factor $>$100 at the gap edge, 
which is consistent with submm constraints from Andrews \etal (2011). In the near future, ALMA can determine how sharp the opacity
decreases in greater detail, which is important to distinguish a pure grain growth scenario (\S 6.6) and scenarios  involving a gaseous gap.

More generally, ALMA can test all the three scenarios above. For gap opening by multiple planets without dust filtration,
 both gas (probed by molecular lines) and dust (probed by dust continuum) 
inside the gap should be equally depleted. Furthermore, in this scenario, the sharpness of the outer gap edge and inner gap edge should 
look similar due to the symmetric nature of the gaseous gap shape. If dust filtration is at work,
the mm sized dust will be more depleted inside the gap than the gas, and the inner gap edge in submm images should
be smoother, or even disappear, than the outer gap edge (Fig. \ref{fig:fig4}). To test whether there
is grain growth inside the gap, we can use multiple wavelength observations to see how the
slope of the opacity changes. However, the dust radial drift may change the big-dust distributions in the inner disk. 

ALMA can also test if the dust and gas have similar density profiles at the outer disk beyond the gap.
Our simulations suggest that, combining dust drift and diffusion, the dust and gas surface density profiles in the outer disk can 
be quite different (\S 5.4).
Although submm observations (Andrews \etal 2012) have indeed suggested that the 
dust disk is more compact than the gaseous disk,
the different surface density slopes between the dust and gas disks predicted in \S 5.4 need to 
be tested by future ALMA observations.
But we note that dust growth and fragmentation can potentially change this relationship 
(e.g. Birnstiel \etal 2012).

\subsection{Other Possibilities}
As illustrated above, the key ingredient to reproduce the GM Aur SED is reducing the abundance of micron sized dust
particles by five orders of magnitude in the inner disk while maintaining a sufficient gas accretion. 
Due to the good coupling between micron sized particles
and the gas, any theory only considering gap formation in a gaseous disk is not enough, such as photoevaporation,
gap opening by planets, etc. Dust growth and settling have to be considered. Besides filtration combined with dust growth,
there are several other possibilities:

The first alternative is purely dust growth. If dust can grow significantly in the inner disk to make the inner disk
optically thin, it can explain GM Aur's SED. In this scenario it may suggest transitional disks are older
than CTTS, which has not been suggested by observations. Furthermore, both SED fitting and sub-mm
observations suggest the gap edge is very sharp, inconsistent with the pure dust growth model. However,
with CARMA, Isella et al.(2012) have suggested LkCa 15 can be explained by the pure dust growth model.
Thus, pure dust growth could still be a possible solution.

The second alternative is a large gas mass reservoir close to the central star with a wide and deep gap beyond.
The wide and deep gap can be caused by a very massive planet(or even a star) or several planets so there is no accretion flow
from the outer disk to the inner disk. The accretion onto the star is sustained by the mass reservoir close
to the planet (e.g. a dead zone). However this mass reservoir needs to be very narrow or depleted in
small-dust to not produce too much near-IR flux.

The third alternative is the dust being held back by the radiation pressure (Chiang 
\& Murray-Clay 2007). However, the efficiency of the radiation pressure is questioned by Dominik 
\& Dullemond (2011).

The fourth alternative is planet gap opening in layered disks. The pros and cons of this model will be discussed in a forthcoming paper.  

However, there is one challenge to all the scenarios trying to explain transitional disks with gap opening by planet(s) located
at a few tens of AU from the central star. 
In the core-accretion scenario, the solid component in the inner disk is likely to have undergone significant growth to form 
planetesimals and planetary embryos within the time taken to form the putative planet at $\sim 20$ AU. 
These planetesimals should collide and continuously regenerate small dust grains. As we have alluded to earlier in this paper, 
one way round this problem is
to invoke the presence of multiple planets in the inner disk to clear these planetesimals. But this hypothesis clearly
requires further investigation. 

\subsection{Transitional VS. Pre-transitional Disks}
Dust filtration also suggests more massive planets can lead to stronger dust filtration and depletion. Thus
transitional disks may have higher mass planet(s) than pre-transitional disks have. Since a higher mass planet
exerts a stronger torque on the outer disk, it may slow down the accretion flow passing the planet 
and lead to a lower disk accretion rate onto the star. This is
consistent with observations that transitional disks have lower accretion rates than pre-transitional 
disks (Espaillat \etal 2012).

Regarding dust growth, transitional disks put strict constraints on the dust abundance in the inner disk since the dust
is optically thin. If these systems have moderate $\dot{M}$, we know they have a significant amount of gas and thus dust to gas ratio needs
to be depleted. On the other hand, for transitional disks having little $\dot{M}$, the disk may harbor a massive companion (e.g. brown dwarf).
In this case, both gas and dust components of the inner disk are significantly reduced below the detection limit and we know little about
their dust to gas ratio, and thus dust growth/depletion is not necessary.

For pre-transitional disks, the dust abundance
in the inner disk is difficult to constrain since it is optically thick. Thus it's possible
to explain pre-transitional disks without invoking dust growth (e.g. dust filtration, multiple-planets, 
photoevaporation Alexander \& Armitage 2007). However it's also possible that the 
planet(s) in pre-transitional disk is/are less massive (making the gap less sharp) so that more 
dust passes through the planet-induced gap and the inner disk remains optically thick even with dust growth. 

\section{Conclusion}

In this paper, we have used two-dimensional two-fluid simulations, a one dimensional model, and
analytic arguments to study dust filtration by the tidally-induced gap outer edge. We have found 
that dust diffusion and the high gas
velocity at the gap edge significantly lower the dust filtration efficiency. Only particles equal or larger
than 0.1 mm can be filtered by a planet-induced gap if the disk has $\alpha$=0.01, $\dot{M}=10^{-8}\msunyr$ and the planet
mass is a few Jupiter masses. These results can
be partly scaled to disks having different mass accretion rates with one dimensionless parameter $T_{s}/\alpha$. With better understanding
of the disk and gap structure, we may be able to constrain the planet mass with future multi-wavelength observations (optical/near-IR scattered images, ALMA etc.).

We have applied this dust filtration threshold (0.1 mm) to transitional disks, and by using a Monte-Carlo radiative transfer model
we have shown that dust filtration alone has difficulties in explaining transitional disk observations, especially for systems with
moderate $\dot{M}$ (e.g. GM Aur). The same difficulty
is suffered by the multiple planets scenario. One possible solution is combining dust filtration with dust growth in the inner disk,
although we have also discussed other possibilities. 
We conclude that dust filtration is a natural consequence of gap opening in protoplanetary disks, and 
although it has some difficulties to explain the near-IR deficit of some transitional disks (which may require additional processes such as dust growth), 
it has important implications for future observations.   

\acknowledgments
This work was supported in part by NASA grant NNX08AI39G from the Origins
of Solar Systems program, and in part by the University of Michigan. Z.Z and R.D.
were also supported by NSF grant AST-0908269 and Princeton University.
C.~E. was supported by the National Science Foundation under Award No. 0901947.
Z.Z. thanks Xuening Bai, Jim Stone, Roman Rafikov, and Steve Lubow, Ruth Murray-Clay for
helpful discussion. Z.Z. also thanks Eugene Chiang for suggesting the layered accretion 
scenario during the International Summer Institute for Modeling in Astrophysics
(ISIMA) in Beijing organized by Pascale Garaud and
Doug Lin. The authors thank the referee for significantly improving this paper. 

\appendix
\section{Two-fluid simulations VS. SFT approximation VS. 1-D models}
We will compare the dust distribution using three different methods in this section.
We expect SFT approximation should give similar results as Two-fluid simulations
for particles smaller than mm, since the dust stopping time for 1 mm particles
 is 1/10th of the hydrodynamic time step and the SFT approximation is valid. 
 
\begin{figure}
\includegraphics[width=0.42\textwidth]{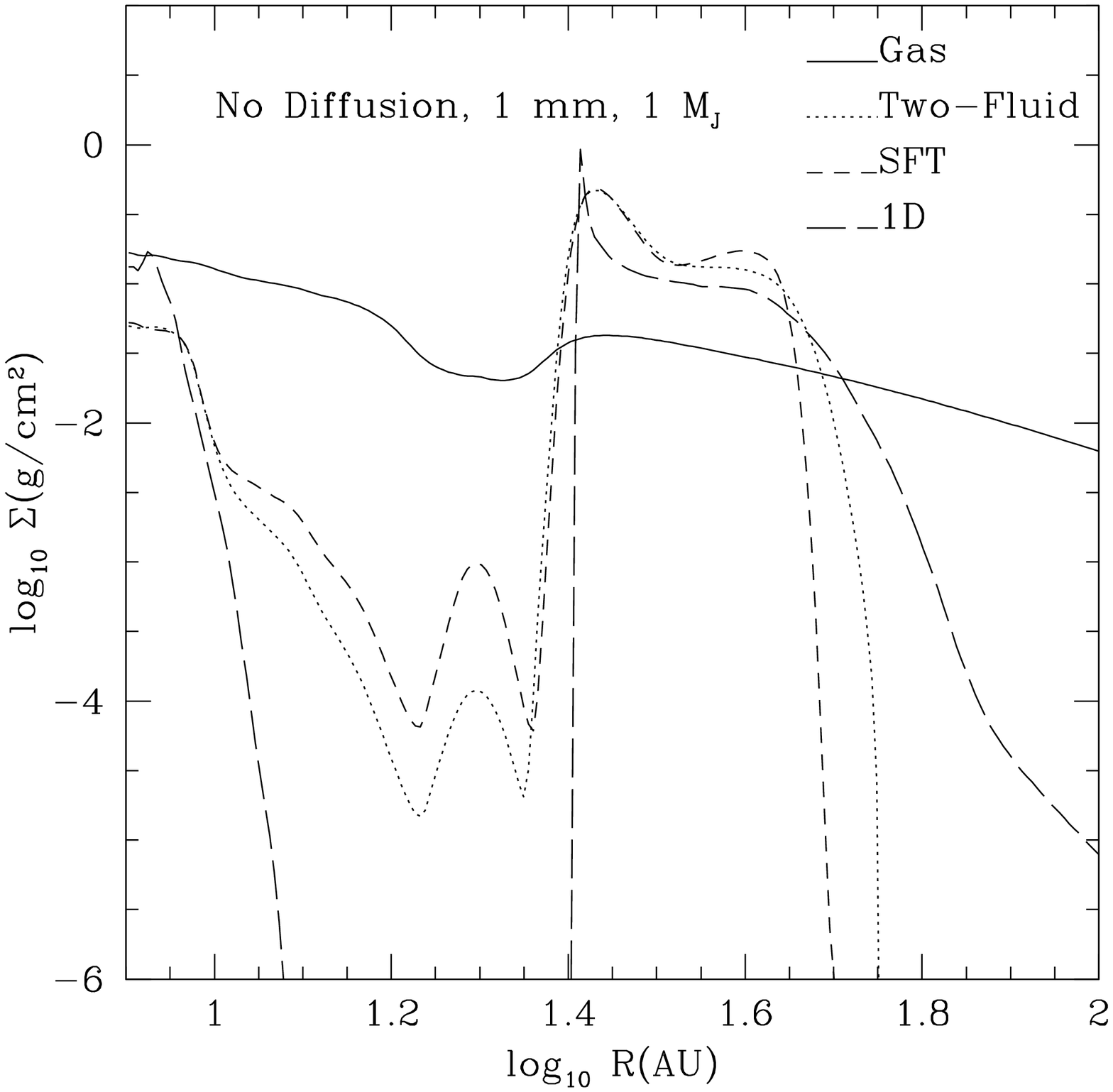} \hfil
\includegraphics[width=0.42\textwidth]{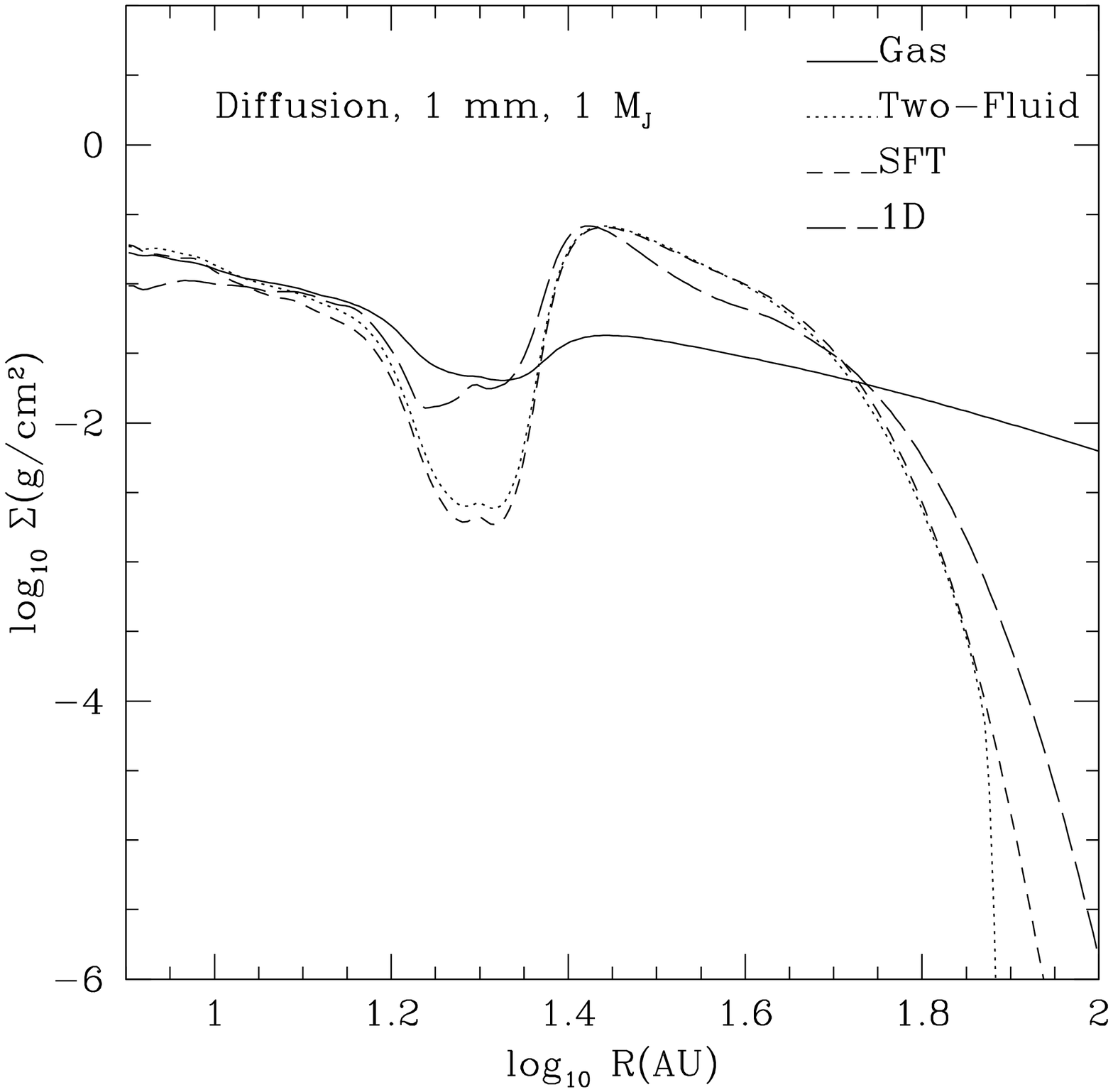} \\
\caption{The azimuthal averaged gas (solid curve) and 1mm dust surface densities 
with(right panel) and without (left panel) dust diffusion at t=5$\times$10$^{4}$ yr. 
The gas surface density is divided by 100 to scale with the dust surface density.
 Results from various methods are compared. The dotted curves are from 2-D two-fluid
 simulations, while the short dashed curves are from the SFT approximation. The long dashed curves
 are from the simplified 1-D simulations. These methods agree with each other outside
 the gap or even at the outer edge of the gap, suggesting they give consistent results
 regarding dust filtration. But 1-D simulations failed to reproduce
 the horseshoe ring around the planet in the left panel.}
\label{fig:fig3}
\end{figure}

The comparison among two-fluid simulations, the SFT approximation, and 1-D models is shown
in Figure \ref{fig:fig3} with (the right panel) and without considering dust diffusion (the left panel).  A 1 M$_{J}$ planet and 1 mm particles
are considered and the simulations have been run to 5$\times$10$^{4}$ yr. 

As shown in Figure \ref{fig:fig3} , the two-fluid simulations and the
SFT approximation agree with each other quite well in both panels. For even smaller particles, 
the assumption of the SFT approximation will be better suited and we would expect the results to be closer to the two-fluid simulations. 

The even simpler 1-D models (the long dashed curves) did a fairly good job at both the inner and outer disk beyond the gap region.
The incapability of 1D models to simulate the horseshoe region is expected since the horseshoe region
has intrinsically a two-dimensional flow pattern and 
material will be trapped inside the horseshoe region.
The flow pattern there is highly non-axisymmetric with the velocity highest around the planet.

In 1-D, with the assumption that the flow is axisymmetric and the radial velocity is given by  Equation \ref{eq:radialgas}, 
material in the horseshoe region
will be quickly  depleted. However, slightly outside the horseshoe region, even at the edge of the gap the flow is still quite
axisymetric. Considering the gap outer edge is where dust filtration takes place,
the 1-D model is still capable to study dust filtration by the gap outer edge,
although the dust surface density at the bottom of the gap is incorrect. 

\FloatBarrier
\clearpage
\begin{table}
\begin{center}
\caption{Models \label{tab1}}
\begin{tabular}{cccccc}

\tableline\tableline
Case name & Method & Diffusion & Planet mass & dust size & Evolution Time  \\
                      &               &        & M$_{J}$  & mm                    & yr \\
\tableline
1 M$_{J}$&&&&&\\
\tableline
1J1mm2F & Two-fluids & No Diff& 1 & 1 & 5$\times$10$^{4}$ yr  \\
1J1mm2FD & Two-fluids & Diff& 1 & 1 & 5$\times$10$^{4}$ yr  \\
1J1mm & SFT approx. & No Diff & 1 & 1 & 1$\times$10$^{5}$ yr  \\
1J0p1mm & SFT approx. & No Diff & 1 & 0.1 & 2.5$\times$10$^{5}$ yr  \\
1J0p03mm & SFT approx. & No Diff & 1 & 0.03 & 5$\times$10$^{5}$ yr  \\
1J1mmD & SFT approx. & Diff & 1 & 1 & 1$\times$10$^{5}$ yr  \\
1J0p1mmD & SFT approx. & Diff & 1 & 0.1 & 2.5$\times$10$^{5}$ yr  \\
1J0p03mmD & SFT approx. & Diff & 1 & 0.03 & 5$\times$10$^{5}$ yr  \\
\tableline
3 M$_{J}$&&&&&\\
\tableline
3J1mm & SFT approx. & No Diff & 3 & 1 & 1$\times$10$^{5}$ yr  \\
3J0p1mm & SFT approx. & No Diff & 3 & 0.1 & 2.5$\times$10$^{5}$ yr  \\
3J0p03mm & SFT approx. & No Diff & 3 & 0.03 & 5$\times$10$^{5}$ yr  \\
3J1mmD & SFT approx. & Diff & 3 & 1 & 1$\times$10$^{5}$ yr  \\
3J0p1mmD & SFT approx. & Diff & 3 & 0.1 & 2.5$\times$10$^{5}$ yr  \\
3J0p03mmD & SFT approx. & Diff & 3 & 0.03 & 5$\times$10$^{5}$ yr  \\
\tableline
6 M$_{J}$&&&&&\\
\tableline
1J1mm & SFT approx. & No Diff & 6 & 1 & 1$\times$10$^{5}$ yr  \\
1J0p1mm & SFT approx. & No Diff & 6 & 0.1 & 2.5$\times$10$^{5}$ yr  \\
1J0p03mm & SFT approx. & No Diff & 6 & 0.03 & 5$\times$10$^{5}$ yr  \\
1J1mmD & SFT approx. & Diff & 6 & 1 & 1$\times$10$^{5}$ yr  \\
1J0p1mmD & SFT approx. & Diff & 6 & 0.1 & 2.5$\times$10$^{5}$ yr  \\
1J0p03mmD & SFT approx. & Diff & 6 & 0.03 & 5$\times$10$^{5}$ yr  \\
\tableline
\end{tabular}
\end{center}
\end{table}

\end{document}